\documentclass[twocolumn,superscriptaddress,showpacs,prl,amsmath,amssymb]{revtex4}
\usepackage{graphicx}
\usepackage{dcolumn}
\usepackage{bm}

\usepackage{color}

\begin{document}

\title{Isoscalar Giant Monopole, Dipole, and Quadrupole Resonances in $^{90,92}$Zr and $^{92}$Mo}
\author{Y. K. Gupta}
\affiliation{Physics Department, University of Notre Dame, Notre Dame, IN 46556}
\affiliation{Nuclear Physics Division, Bhabha Atomic Research Centre, Mumbai, 400085, India}
\author{K. B. Howard}
\affiliation{Physics Department, University of Notre Dame, Notre Dame, IN 46556}
\author{U. Garg}
\affiliation{Physics Department, University of Notre Dame, Notre Dame, IN 46556}
\author{J. T. Matta}
\thanks{Present address: Physics Division, Oak Ridge National Laboratory, Oak Ridge, TN 37380}
\affiliation{Physics Department, University of Notre Dame, Notre Dame, IN 46556}
\author{M.~{\c S}enyi{\u g}it}
\affiliation{Physics Department, University of Notre Dame, Notre Dame, IN 46556}
\affiliation{Department of Physics, Faculty of Science, Ankara University, TR-06100 Tando{\u g}an, Ankara, Turkey}

\author{M. Itoh} 
\affiliation{Cyclotron and Radioisotope Center, Tohoku University, Sendai 980-8578, Japan}
\author{S. Ando} 
\affiliation{Cyclotron and Radioisotope Center, Tohoku University, Sendai 980-8578, Japan}
\author{T. Aoki} 
\affiliation{Cyclotron and Radioisotope Center, Tohoku University, Sendai 980-8578, Japan}
\author{A. Uchiyama} 
\affiliation{Cyclotron and Radioisotope Center, Tohoku University, Sendai 980-8578, Japan}

\author{S. Adachi}
\affiliation{Research Center for Nuclear Physics (RCNP), Osaka University, Osaka 567-0047, Japan}
\author{M. Fujiwara}
\affiliation{Research Center for Nuclear Physics (RCNP), Osaka University, Osaka 567-0047, Japan}
\author{C. Iwamoto}
\affiliation{Research Center for Nuclear Physics (RCNP), Osaka University, Osaka 567-0047, Japan}
\author{A. Tamii}
\affiliation{Research Center for Nuclear Physics (RCNP), Osaka University, Osaka 567-0047, Japan}

\author{H. Akimune}
\affiliation{Department of Physics, Konan University, Hyogo 658-8501, Japan}
\author{C. Kadono}
\affiliation{Department of Physics, Konan University, Hyogo 658-8501, Japan}
\author{Y. Matsuda}
\affiliation{Department of Physics, Konan University, Hyogo 658-8501, Japan}
\author{T. Nakahara}
\affiliation{Department of Physics, Konan University, Hyogo 658-8501, Japan}

\author{T. Furuno}
\affiliation{Department of Physics, Kyoto University, Kyoto 606-8502, Japan}
\author{T. Kawabata}
\affiliation{Department of Physics, Kyoto University, Kyoto 606-8502, Japan}
\author{M. Tsumura}
\affiliation{Department of Physics, Kyoto University, Kyoto 606-8502, Japan}

\author{M. N. Harakeh}
\affiliation{KVI-CART, University of Groningen, 9747 AA Groningen, The Netherlands}
\author{N.  Kalantar-Nayestanaki }
\affiliation{KVI-CART, University of Groningen, 9747 AA Groningen, The Netherlands}

\date{\today}

\begin{abstract}
The isoscalar giant monopole, dipole, and quadrupole strength distributions have been deduced  in $^{90, 92}$Zr, and $^{92}$Mo from ``background-free'' spectra of inelastic $\alpha$-particle scattering at a beam energy of 385 MeV at extremely forward angles, including 0$^{\circ}$. These strength distributions were extracted by a multipole-decomposition analysis based on the expected angular distributions of the respective multipoles. 
All these strength distributions for the three nuclei practically coincide with each other, affirming that giant resonances, being collective phenomena, are not influenced by nuclear shell structure near $A\sim$90, contrary to the claim in a recent measurement. 
\end{abstract}


\maketitle

\section{\label{sec:level1}Introduction}
Giant resonances are high-frequency fundamental modes of nuclear collective excitation. In particular, the isoscalar giant monopole (ISGMR) and dipole (ISGDR) resonances are compressional modes of nuclear density oscillation of the finite nuclear systems. Direct experimental information on the nuclear incompressibility of infinite nuclear matter, $K_{\infty }$, can be obtained only from these ``compressional mode" oscillations of finite nuclei. Nuclear incompressibility characterizes the nuclear equation of state (EOS) which in turn provides crucial information towards the understanding of wide-ranging phenomena  such as the radii of neutron stars, the strength of supernova explosions, transverse flow in relativistic heavy-ion collisions, and the nuclear skin thickness \cite{Harakeh_book, Lattimer2001}. The centroid energies of the compressional modes, $E_\mathrm{ISGMR}$ and $E_\mathrm{ISGDR}$, are directly related to the nuclear incompressibility of the finite nucleus; in the scaling model, these relationships are expressed as \cite{str1982, Treiner1981}:

\begin{equation}
E_\mathrm{ISGMR}=\hbar\sqrt{\frac{K_{A}}{m\left<r^{2}\right>_{0}}},
\label{EISGMR}
\end{equation}

\begin{equation}
E_\mathrm{ISGDR}=\hbar\sqrt{\frac{7K_{A}+\frac{27}{25}\epsilon_{F}}{3m\left<r^{2}\right>_{0}}},
\label{EISGDR}
\end{equation}

\noindent 
where $K_{A}$ is the incompressibility of a finite nucleus with mass number $A$,  $\left<r^{2}\right>_{0}$ is the ground-state mean square radius of the nucleus, $m$ is the nucleon mass, and $\epsilon_{F}$ is the Fermi energy \cite{str1982}. The determination of $K_{\infty }$ from $K_{A}$ is achieved within a framework of self-consistent RPA calculations, using the widely accepted method described by Blaizot \cite{Blaizot1995}. Because the compressional modes are collective phenomena, the determination of $K_{\infty }$ is believed to be independent of the choice of the nucleus, provided that approximately 100\% of the energy weighted sum rule (EWSR) is exhausted in a single giant resonance peak; this condition is satisfied for sufficiently heavy nuclei ($A\geq 90$) \cite{Harakeh_book}. The presently accepted value of $K_{\infty}$, determined from ISGMR in ``standard" nuclei such as $^{90}$Zr and $^{208}$Pb, is 240 $\pm$ 20 MeV \cite{Colo_Giai_PRC2004, Todd-Rutel_PRL2005, colo_garg_sagawa, pickarewickz2014}. 

Experimental determination of $E_\mathrm{ISGMR}$ and $E_\mathrm{ISGDR}$ is not straightforward primarily due to the overlap with the isoscalar giant quadrupole resonance (ISGQR) and the isoscalar high-energy octupole resonance (ISHEOR), respectively. The ``bimodal strength distribution" of the ISGDR further complicates this situation \cite{Uchida_90Zr, Uchida_PLB2003, Itoh_prc2003, nayak2006}. Nevertheless, with data of high quality and careful analyses, it has been shown that the $K_{A}$ determined from ISGDR is consistent with that from ISGMR \cite{Uchida_90Zr, Uchida_PLB2003}.

Study of  these ``compressional modes"  has been carried out for a variety of nuclei during the last two decades. 
One of the major thrusts of these studies in recent years has been to determine the asymmetry component of the nuclear incompressibility, $K_{\tau}$. ISGMR data in the isotopic chains of Sn and Cd have yielded a consistent value of $K_{\tau}$ = -550 $\pm$ 100 MeV. Most intriguingly, Sn and Cd isotopes have been observed to be ``soft" in sharp contrast to the ``standard" doubly magic nuclei, and the question,``why are the Sn and Cd isotopes so soft?" remains unanswered. It has been cited as one of the open problems in nuclear structure physics \cite{fluffy1,Li_PRL2007, Li_2010, Darshana2012,fluffy2,Speth2009,fluffy6,fluffy5,fluffy4}. Experimentally determined $K_{A }$ values for these ``soft" elements,  however, vary quite smoothly over the wide isotopic chains \cite{Li_PRL2007, Li_2010, Darshana2012}.

Recently, the Texas A \& M group has reported isoscalar giant resonance  strength distributions for $L\leq3$ in several isotopes of Zr and Mo \cite{YB_ZrMo2013, YB_Mo2015, Krishi_2015, Button2016}. Their results were rather unexpected in that the extracted ISGMR strengths varied in a very dramatic manner in these nuclei. In particular, the $A$=92 nuclei, $^{92}$Zr and $^{92}$Mo, emerged quite disparate from the others: the ISGMR energies ($E_\mathrm{ISGMR}$) for $^{92}$Zr and $^{92}$Mo were observed to be, respectively, 1.22 and  2.80 MeV higher than that of $^{90}$Zr.  Consequently, the $K_{A}$ values determined for $^{92}$Zr and $^{92}$Mo were, respectively, $\sim$27 MeV and $\sim $56 MeV higher than the $K_{A}$ for $^{90}$Zr. These results, if correct, imply significant nuclear structure contribution to the nuclear incompressibility in this mass region. Such nuclear structure effects have not been observed in any of the investigations of ISGMR going back to its first identification  in the late 1970's  \cite{Harakesh_prl1977, YBPRL1977} and, indeed, would be contrary to the standard hydrodynamical picture associated with this mode of collective oscillation \cite{LIPPARINI1989}.

Very recently, we have shown unambiguously that ISGMR response in the $^{90, 92}$Zr and $^{92}$Mo nuclei  is virtually identical \cite{YKGPLB2016}.  In the present article, we report on detailed investigations for the isoscalar giant monopole, dipole and quadrupole resonances (ISGMR, ISGDR and ISGQR)  in the $^{90, 92}$Zr and $^{92}$Mo nuclei from inelastic $\alpha$-scattering measurements. It is shown that not only the ISGMR, but also the other ``compressional mode", ISGDR, and the ISGQR response as well are almost identical in these nuclei, revealing no influence of open and/or closed shells for protons and/or neutrons on the collective modes of nuclear excitations.

\section{\label{sec:level2}Experimental Procedures}

Inelastic scattering of 385-MeV $\alpha $ particles
was measured at the ring cyclotron facility
of the Research Center for Nuclear Physics (RCNP), Osaka
University. Self-supporting  foils of highly enriched  targets ($97.70\%$, $95.13\%$, and $97.37\%$ for $^{90}$Zr, $^{92}$Zr, and $^{92}$Mo, respectively)   were  used, with thicknesses ranging from 4.0 to 5.4 mg/cm$^{2}$. Inelastically scattered $\alpha$ particles were momentum analyzed with the high-resolution magnetic spectrometer ``Grand
Raiden" \cite{Fujiwara_GR}, and their horizontal and vertical positions were measured with a focal-plane detector
system composed of two position-sensitive multi-wire drift
chambers (MWDCs) and two plastic scintillators \cite{Itoh_prc2003}. These detectors enabled particle identification and
reconstruction of the trajectories of scattered particles.  

\begin{figure}
\centering\includegraphics [trim= 0.11mm 0.5mm 0.1mm 0.1mm,
angle=360, clip, height=0.25\textheight]{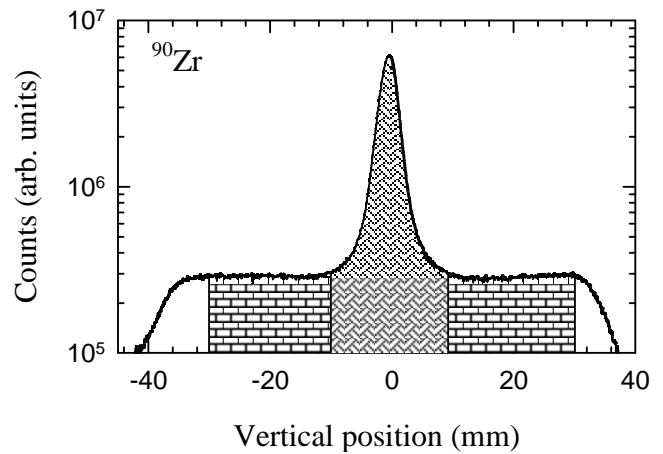}
\caption{A vertical position spectrum  measured at the spectrometer angle of $\theta$=0$^{\circ}$ (average angle 0.7$^{\circ}$ in the laboratory frame).  The central region represents true+instrumental background events. The off-center regions represent only instrumental background events. The true events were obtained by subtracting instrumental background events from the true+instrumental background events. }
\label{ZeroDegYc}
\end{figure}

\begin{figure} 
\centering\includegraphics [trim= 0.11mm 0.5mm 0.1mm 0.1mm,
angle=360, clip, height=0.42\textheight]{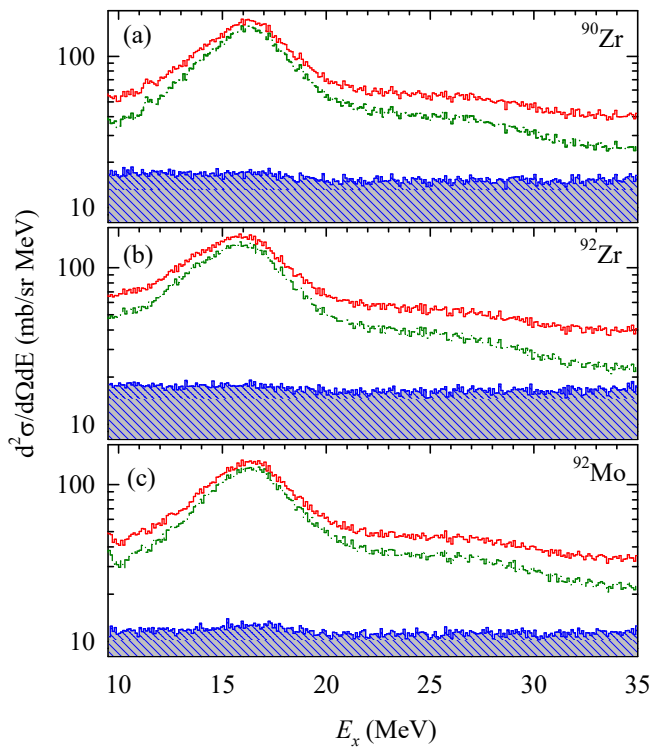}
\caption{(Color online) Excitation-energy spectra for the ($\alpha, \alpha'$) reaction at $E_{\alpha}$=385 MeV at the spectrometer angle of $\theta$=0$^{\circ}$ (average angle 0.7$^{\circ}$ in the laboratory frame) for $^{90}$Zr, $^{92}$Zr, and $^{92}$Mo in panels (a), (b), and (c), respectively. In the each panel, the blue hatched region represents the instrumental background. The solid red and green histograms show the energy spectra before and after the instrumental-background subtraction, respectively.}
\label{ZeroDeg}
\end{figure}

The vertical-position spectrum obtained in the double-focusing
mode of the spectrometer was exploited to eliminate the
instrumental background \cite{Uchida_PLB2003,Itoh_prc2003}.
A typical vertical-position spectrum measured at the spectrometer angle of $\theta$=0$^{\circ}$ for $^{90}$Zr is shown in Fig. \ref{ZeroDegYc}, where the central region represents true+background (instrumental) events and the off-center regions represent only background (instrumental) events. The background shapes are almost identical on both sides of the true+background peak. The true events were obtained by subtracting background events from the true+background events. Figs. \ref{ZeroDeg}[(a)--(c)] show the instrumental background, and excitation-energy spectra before and after the background subtraction, as measured at the spectrometer angle $\theta$=0$^{\circ}$ for the three nuclei. One sees in Fig. \ref{ZeroDeg} that the instrumental background is almost constant in the entire excitation energy range. 
The instrumental background is observed to be maximum in the 0$^{\circ}$ measurement and reduces quite significantly at higher angles (except at 6$^{\circ}$, where it increases somewhat because of the presence of a Faraday cup inside the scattering chamber). The behavior of instrumental background is almost identical for all three nuclei  considered in the present work. 

Inelastic scattering measurements were performed  at very forward central angles of the spectrometer (from 0$^{\circ }$ to 9.5$^{\circ }$) and at magnetic-field settings corresponding to excitation energies in the range $E_x =$ 9.5--32.5 MeV. The scattering angles were averaged over the acceptance of Grand Raiden. Ray-tracing technique was used to reconstruct the horizontal scattering angle and the effective angular width of 1.6$^{\circ }$ for each central angle was divided into four equal regions in the offline data analysis. Thus, measurements at one angle setting of the Grand Raiden provided four data points. Data were also taken with a $^{12}$C target at each setting, providing a precise energy calibration. Energy losses in the target foils for the incident beam and outgoing $\alpha$ particles were duly taken into account.   

\begin{figure}[h] 
\centering\includegraphics [trim= 0.11mm 0.5mm 0.1mm 0.1mm,
angle=360, clip, height=0.32\textheight]{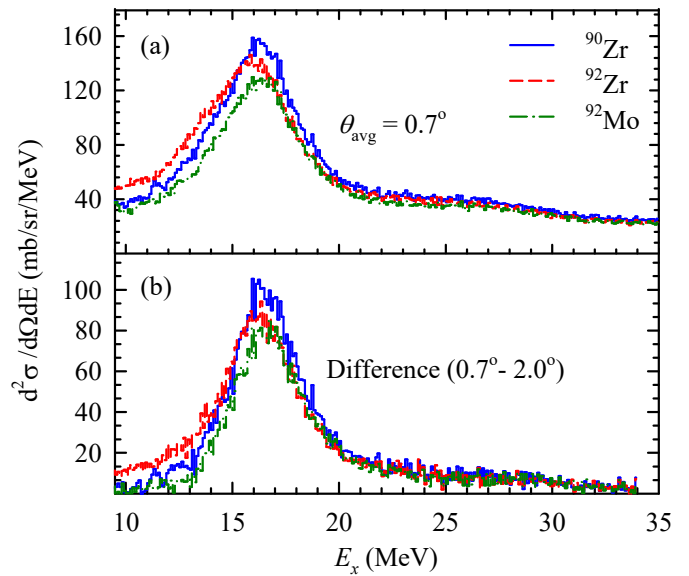}
\caption{ (Color online) (a) Excitation-energy spectra measured at the spectrometer angle of $\theta$=0$^{\circ}$ (average angle 0.7$^{\circ}$ in the laboratory frame) for the three nuclei, $^{90}$Zr (solid blue), $^{92}$Zr (red dashed), and $^{92}$Mo (green dash-dotted). (b) Difference spectra of average angles 0.7$^{\circ}$ and 2.0$^{\circ}$ for the three nuclei, $^{90}$Zr (solid blue), $^{92}$Zr (red dashed), and $^{92}$Mo (green dash-dotted). The difference spectra comprise essentially  the monopole strength (see text).}
\label{ZeroDeg_DiffSpect}
\end{figure}

\section{\label{sec:level3}Data Analysis}
The excitation-energy spectra at the spectrometer angle of $\theta$=0$^{\circ}$ (average angle 0.7$^{\circ}$ in the laboratory frame) for the three nuclei are overlaid in Fig. \ref {ZeroDeg_DiffSpect}(a). The spectra near 0$^{\circ}$ scattering angle exhibit predominantly the monopole strength, and the 0.7$^{\circ}$ spectra for the three nuclei, shown in Fig. \ref {ZeroDeg_DiffSpect}(a), are very similar; in particular, for excitation energies beyond 20 MeV, these are nearly identical whereas the results in Refs. \cite{YB_ZrMo2013, YB_Mo2015, Krishi_2015} had shown marked differences in the extracted ISGMR strength in this excitation-energy region. The minor differences in the low-energy part of the spectra  (below 16 MeV) are primarily  due to the different shapes of the ISGMR at low energy and could also be partly due to the different contributions from the non-compressional L=1 strength, as discussed later.

The ``difference-spectrum", obtained from subtracting the inelastic scattering spectrum at the first minimum of the expected ISGMR angular distribution from that at 0$^{\circ}$ (where the ISGMR strength is maximal), essentially represents only the ISGMR strength. This is a consequence of the fact that all other multipolarities have relatively flat distributions in this angular region and, thus, are subtracted out in the ``difference-spectrum" (see, Ref. \cite{BRANDENBURG1987}). The difference spectra for average angles of 0.7$^{\circ}$ (maximal ISGMR strength) and 2.0$^{\circ}$ (first minimum of ISGMR strength) for all the three nuclei are also almost identical, as shown in Fig. \ref {ZeroDeg_DiffSpect}(b), again indicating similar ISGMR response in the three nuclei. In particular, the difference spectra beyond $E_x$=20 MeV fully coincide with each other, again quite different from the extracted ISGMR strengths in Refs. \cite{YB_ZrMo2013, YB_Mo2015, Krishi_2015}.

\begin{figure*}[t]
\centering\includegraphics [trim= 0.5mm 0.5mm 0.1mm 0.1mm,
angle=360, clip, height=0.55\textheight]{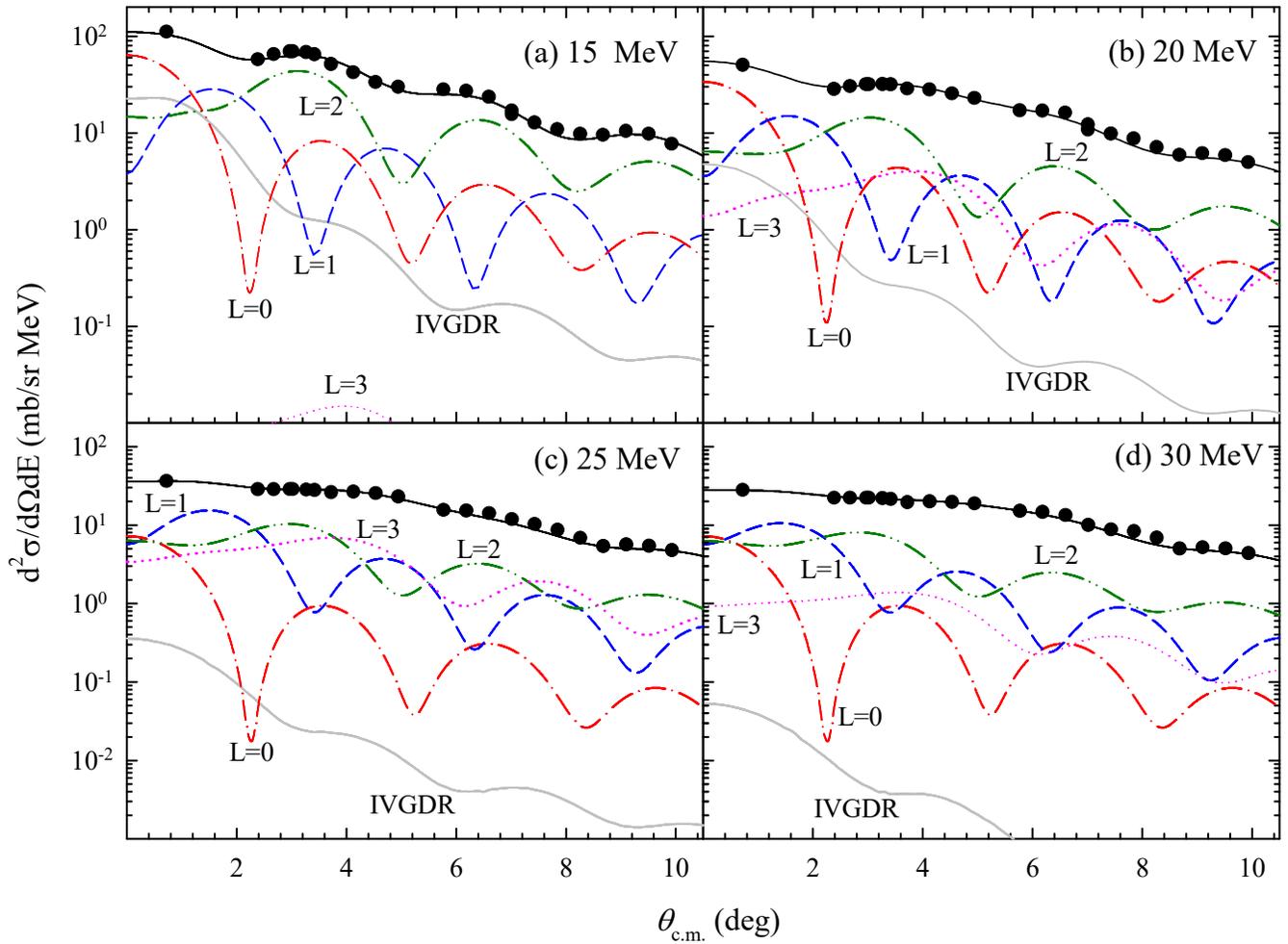}
\caption{(Color online) Typical angular distributions for inelastic $\alpha$
scattering from $^{90}$Zr. The solid line (black) through the data shows the sum of various multipole
components obtained from MDA. The dash-dotted (red), dashed (blue), dash-double-dotted (green), and dotted (magenta) curves
indicate contributions from L = 0, 1, 2, and 3, respectively, with the transferred angular momentum L indicated for each of the curves. The solid gray line shows the IVGDR contribution.}
\label{MDA_90Zr}
\end{figure*}

\begin{figure*}[t]
\centering\includegraphics [trim= 0.5mm 0.5mm 0.1mm 0.1mm,
angle=360, clip, height=0.55\textheight]{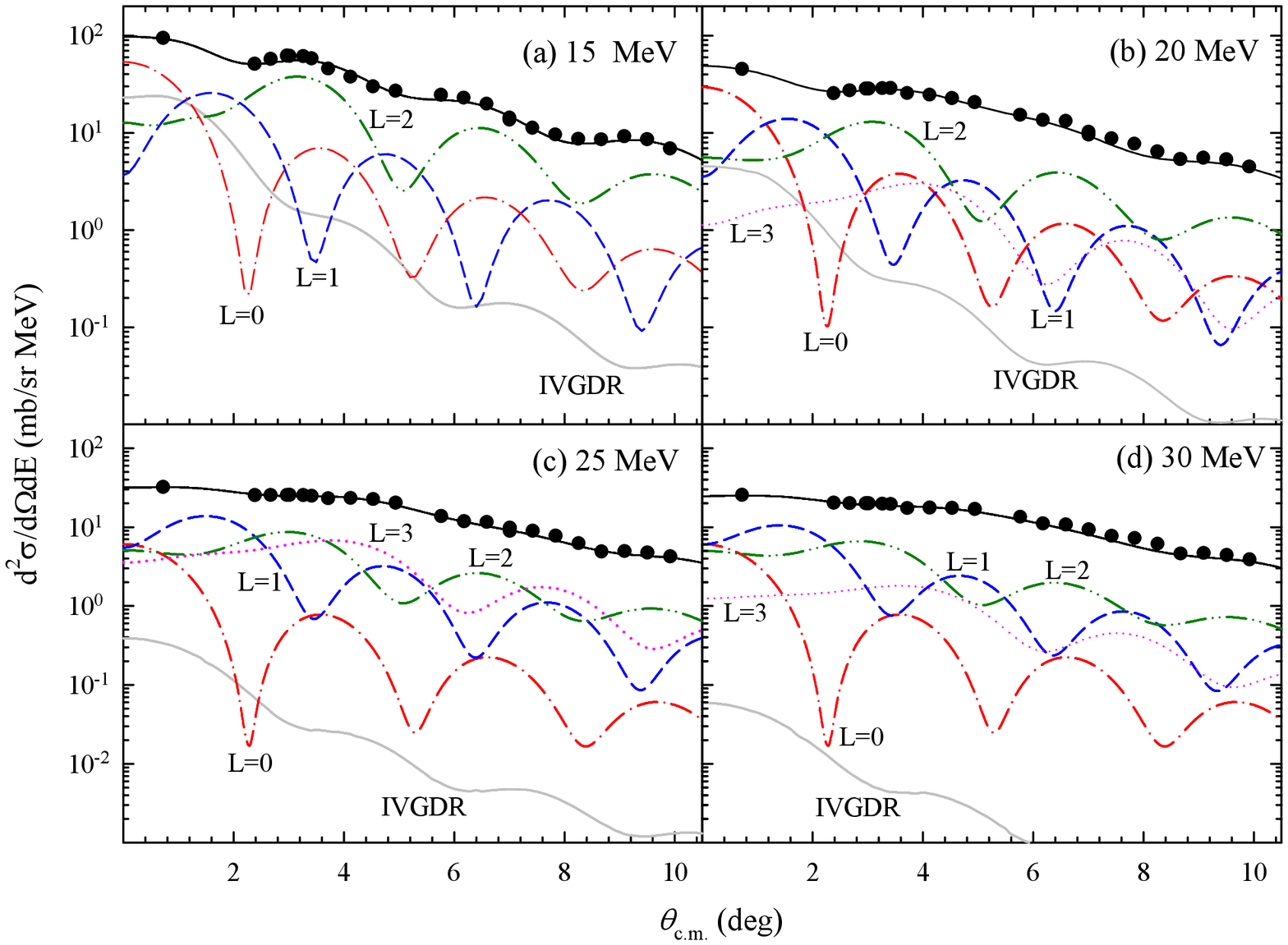}
\caption{(Color online) Typical angular distributions for inelastic $\alpha$
scattering from $^{92}$Mo. The solid line (black) through the data shows the sum of various multipole
components obtained from MDA. The dash-dotted (red), dashed (blue), dash-double-dotted (green), and dotted (magenta) curves
indicate contributions from L = 0, 1, 2, and 3, respectively, with the transferred angular momentum L indicated for each of the curves. The solid gray line shows the IVGDR contribution.}
\label{MDA_92Mo}
\end{figure*}

We have employed the standard multipole-decomposition analysis (MDA) procedure \cite{Li_2010, YKG_PLB2015} to extract the giant resonance strengths for different multipolarities. Experimental cross-sections were binned into 1-MeV intervals. The laboratory angular distributions for each excitation-energy bin were converted to the center-of-mass frame using the standard Jacobian and relativistic kinematics. Typical angular distributions at excitation energies of 15, 20, 25, and 30 MeV for $^{90}$Zr and $^{92}$Mo are presented in Figs. \ref{MDA_90Zr} and \ref{MDA_92Mo}, respectively.  The experimental double-differential cross sections are
expressed as linear combinations of calculated double-differential cross sections associated with different
multipoles as follows \cite{Li_2010, YKG_PLB2015}:
\begin{equation}
\frac{d^{2}\sigma ^{\mathrm{exp}} (\theta_{\mathrm{c.m.}}, E_{x})}{d\Omega dE}= \sum\limits_{L=0}^8 a_{L}(E_{x})\frac{d^{2}\sigma_{L}^{\mathrm{DWBA}}(\theta_{\mathrm{c.m.}}, E_{x}) }{d\Omega dE} 
\label{MDA}
\end{equation}
\noindent
where $a_{L}(E_{x})$ is EWSR fraction for the L$^{th}$ component and $\frac{d^{2}\sigma_{L}^{\mathrm{DWBA}} }{d\Omega dE} (\theta_{\mathrm{c.m.}}, E_{x})$ is the calculated DWBA cross section corresponding to 100\% EWSR for the L$^{th}$ multipole at excitation energy $E_{x}$. The isovector giant dipole resonance (IVGDR) contribution
was subtracted out of the experimental data prior to the fitting procedure \cite{Darshana2012, Li_2010}. Photonuclear cross-section data  \cite{Berman_1975} were used in conjunction with DWBA calculations based on the Goldhaber-Teller model to estimate the IVGDR differential cross section as a function of scattering angle \cite{Satchler1987}. Lorentzian parameters for the photonuclear cross sections [peak cross section ($\sigma_{m}$), peak energy ($E^\mathrm{photo}_{m}$), and width ($\Gamma^\mathrm{photo}$)] used in the present work were taken from Ref. \cite{Berman_1975}, and are presented in Table \ref{tab:photonuclear}.

\begin{table}[h]
\caption{\label{tab:photonuclear} Lorentzian parameters (from Ref. \cite{Berman_1975}) for the photonuclear cross sections.}

\begin{ruledtabular}
\begin{tabular}{cccc}
Nucleus     &  $\sigma_{m}$   & $E^\mathrm{photo}_{m}$    & $\Gamma^\mathrm{photo}$  \\
            & (mb)            & (MeV)              &(MeV)   \\
\hline\\
$^{90}$Zr   &185              &16.85               &4.02  \\
$^{92}$Zr   &184              &16.58               &4.20\\
$^{92}$Mo   &162              &16.82               &4.14
\end{tabular}
\end{ruledtabular}
\end{table}

In order to determine the optical-model parameters (OMPs) for the DWBA calculations, data for elastic scattering and inelastic scattering to 2$^{+}$ and 3$^{-}$ states were taken for each nucleus in the angular range of 5.0$^{\circ }$ to 26.5$^{\circ }$. The
``hybrid" optical-model potential (OMP) proposed by Satchler and Khoa \cite{Satchler_Khoa1997} was employed. In this procedure, the real part of the OMP is generated by single folding with a density-dependent Gaussian $\alpha$-nucleon interaction \cite{Li_2010}. A Woods-Saxon potential is used for the imaginary term of the OMP. Therefore, the total  $\alpha$-nucleus ground-state potential, $U(r)$, is given 
by:
\begin{equation}
U(r)=-V(r)-\it{i}W/\{1+\exp[(r-R_{I})/a_{I}]\}
\label{OpticalPot}
\end{equation}
where $V(r)$ is the real single-folding potential obtained using the computer code SDOLFIN \cite{DOLFIN} by
folding the ground-state density with the density-dependent
$\alpha $-nucleon interaction:
\begin{equation}
\mathrm{\upsilon_{DDG}}(\mathbf{r},\mathbf{r'},\rho)=-\upsilon[1-\beta \rho(\mathbf{r'})^{2/3}]\mathrm{exp}\left(-\frac{\mathbf{|r-r'|}^2}{t^2}\right). 
\end{equation}

\noindent
Here, $\mathrm{\upsilon_{DDG}}(\mathbf{r},\mathbf{r'},\rho)$ is the density dependent $\alpha$-nucleon interaction, $\mathbf{|r-r'|}$ is the distance between center-of-mass of the $\alpha$-particle and a target nucleon, $\rho(\mathbf{r'})$ is the ground-state density of the target nucleus at a position $\mathbf{r'}$ of the target nucleon, $\beta$=1.9 fm$^{2}$, and $t$=1.88 fm. In Eq.~(\ref{OpticalPot}), $W$ is the depth of the Woods-Saxon type imaginary part of the potential, with the radius $R_{I}$ and diffuseness $a_{I}$. 

\begin{figure}[h] 
\centering\includegraphics [trim= 0.11mm 0.5mm 0.1mm 0.1mm,
angle=360, clip, height=0.52\textheight]{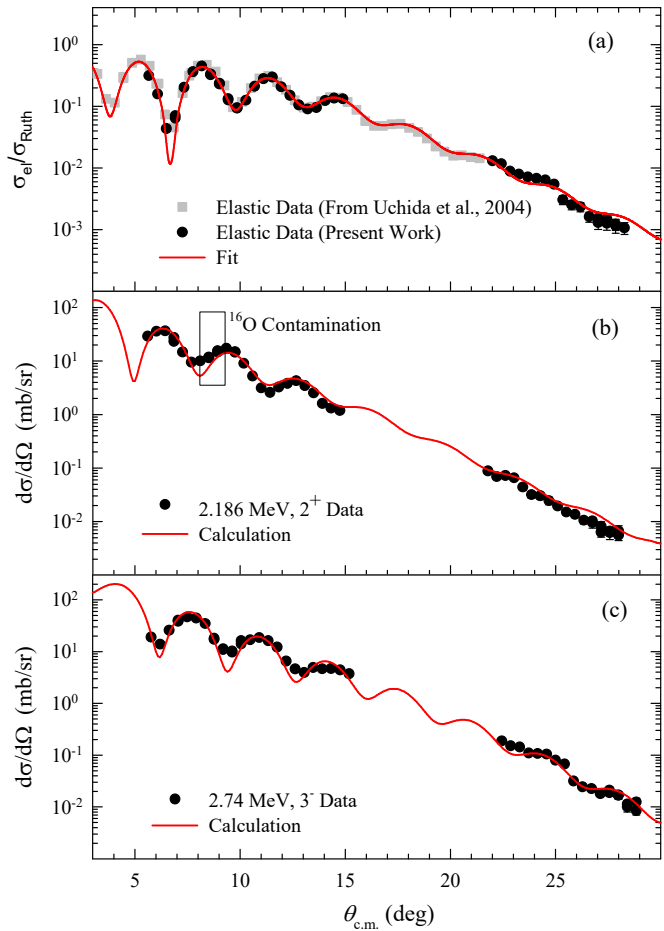}
\caption{(Color online) (a) Angular distribution of the ratio of the differential
cross sections for elastic scattering to Rutherford scattering for
385 MeV $\alpha$ particles from $^{90}$Zr (solid black circles). Solid gray squares are from earlier work \cite{Uchida_90Zr} and the solid red line is the optical-model fit to the data. In panels (b) and (c), angular distribution of differential cross sections for the 2$^{+}$ state and 3$^{-}$ states are shown, where the solid red lines 
show the corresponding results of the DWBA calculations (see text). The rectangular box in panel (b) represents the $^{16}$O contamination region in the $^{90}$Zr target. }
\label{EL_2p_3m_90Zr}
\end{figure}

\begin{figure}[h] 
\centering\includegraphics [trim= 0.11mm 0.5mm 0.1mm 0.1mm,
angle=360, clip, height=0.52\textheight]{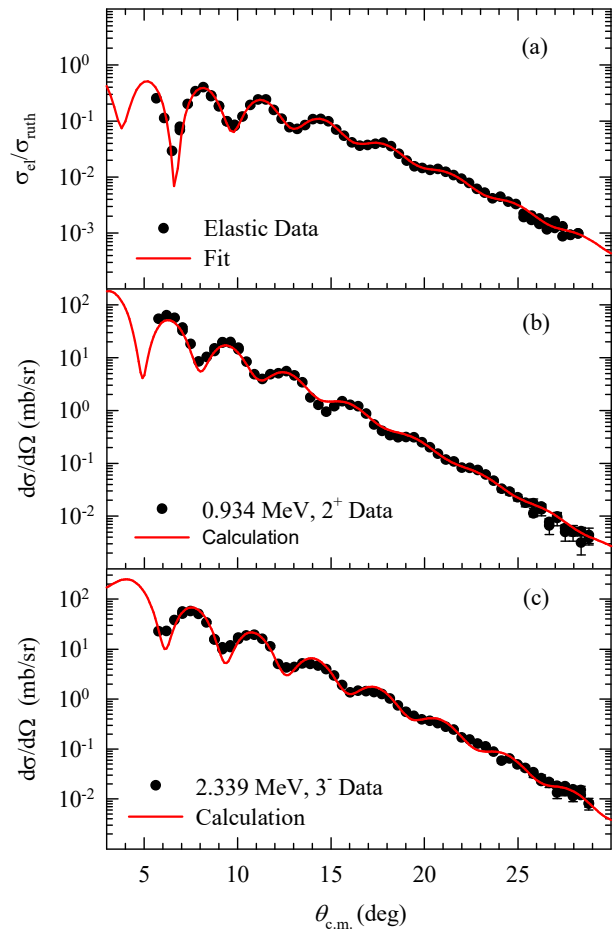}
\caption{(Color online) (a) Angular distribution of the ratio of the differential
cross sections for elastic scattering to Rutherford scattering for
385 MeV $\alpha$ particles from $^{92}$Zr (solid black circles). The solid red line is the optical-model fit to the data. In panels (b) and (c), angular distribution of differential
cross sections for the  2$^{+}$ state and 3$^{-}$ states are shown, where the solid red lines 
show the corresponding results of the DWBA calculations (see text).}
\label{EL_2p_3m_92Zr}
\end{figure}

\begin{figure}[h] 
\centering\includegraphics [trim= 0.11mm 0.5mm 0.1mm 0.1mm,
angle=360, clip, height=0.55\textheight]{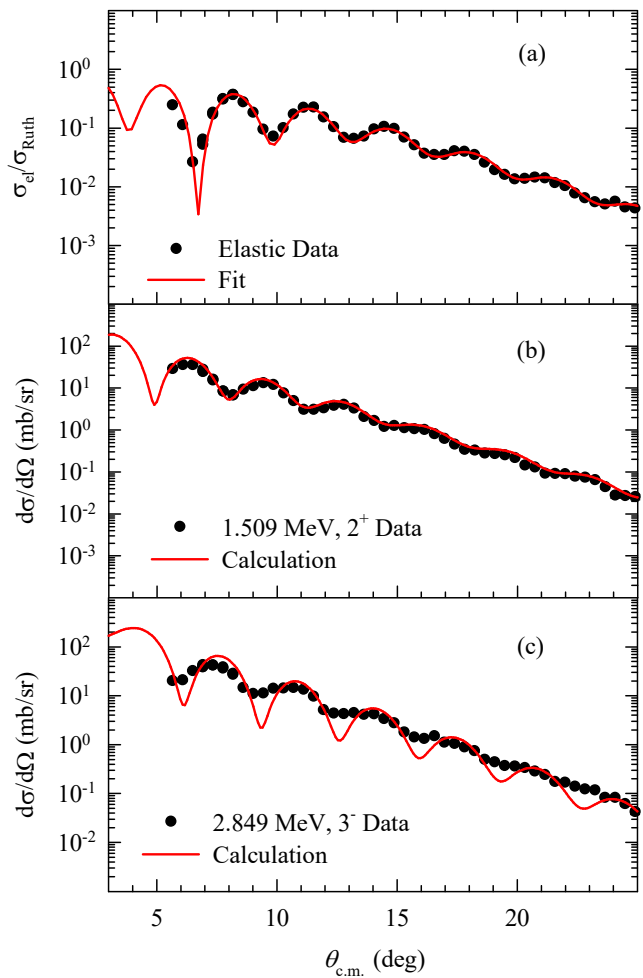}
\caption{Same as Fig. \ref{EL_2p_3m_92Zr}, but for $^{92}$Mo.}
\label{EL_2p_3m_92Mo}
\end{figure}

The imaginary potential parameters ($W$, $R_{I}$, and $a_{I}$), together with the depth of the real part, $V$, were obtained for each nucleus by fitting the elastic-scattering cross sections using the computer code PTOLEMY \cite{ptolemy1,ptolemy2}. Radial moments were obtained by numerical integration of
the Fermi mass distribution using the parameters $c$ and $a$ taken from Ref. \cite{Fricke1995} and given in Table \ref{OMPs}. The best fits to  elastic scattering cross-section data (normalized to the Rutherford cross section) for $^{90,92}$Zr and $^{92}$Mo, obtained from minimization of $\chi^{2}$, are  shown in Figs. \ref{EL_2p_3m_90Zr}(a), \ref{EL_2p_3m_92Zr}(a), and \ref{EL_2p_3m_92Mo}(a), respectively. 
Differential cross sections for the excited states, 2$^{+}$ and 3$^{-}$ are shown, respectively, in the panels (b) and (c) of Figs. \ref{EL_2p_3m_90Zr}, \ref{EL_2p_3m_92Zr}, and \ref{EL_2p_3m_92Mo} for $^{90}$Zr, $^{92}$Zr, and $^{92}$Mo, respectively. Elastic scattering data for $^{90}$Zr were not measured over the full angular range of 5.0$^{\circ }$ to 26.5$^{\circ }$ primarily because elastic scattering  measurements for  $^{90}$Zr have been performed earlier in the angular range of 2.5$^{\circ }$ to 22.5$^{\circ }$ \cite{Uchida_90Zr}. Elastic scattering data from the present work for $^{90}$Zr join smoothly with those measured earlier \cite{Uchida_90Zr} as shown in Fig. \ref{EL_2p_3m_90Zr}(a). 

The optical-model parameters obtained for each nucleus are presented in Table \ref{OMPs}. The charge radii ($R_{C}$), excitation energies and transition probabilities (from literature \cite{BE2_24Mg, BE3}) for the 2$^{+}$ and 3$^{-}$  states used in the present DWBA calculation are also shown in Table \ref{OMPs}. Using these $B(E2)$ and $B(E3)$ values, and the OMPs thus obtained, the angular distributions for the 2$^{+}$ and 3$^{-}$ states for each nucleus were calculated within the  same DWBA framework.  Good agreement between the calculated and experimental angular distributions  for the 2$^{+}$ and 3$^{-}$ states, as shown in panels (b) and (c) of Figs. \ref{EL_2p_3m_90Zr}, \ref{EL_2p_3m_92Zr}, and \ref{EL_2p_3m_92Mo} for $^{90}$Zr, $^{92}$Zr, and $^{92}$Mo, respectively, establishes the appropriateness of the OMPs.

\begin{table*}
\caption{\label{OMPs} Fermi-distribution parameters ($c$ and $a$) from Ref. \cite{Fricke1995} and optical-model parameters obtained by fitting the elastic scattering
data. Also listed are the excitation energies of the low-lying states (2$^+$, 3$^-$) and the corresponding $B(E\lambda )$ values from Refs \cite{BE2_24Mg, BE3}.}
\begin{ruledtabular}
\begin{tabular}{cccccccccc}
Nucleus     &  $c$    & $a$     & V     & W        & $R_{I}$      &$a_{I}$     &$R_{C}$    &[E$_{x}$(2$^+$), $B(E2)$] & [E$_{x}$(3$^-$), $B(E3)$] \\
            & (fm)    & (fm)    &(MeV)  &(MeV)     &  (fm)        & (fm)       &(fm)         & (MeV, e$^{2}$b$^{2}$) & (MeV, e$^{2}$b$^{3}$) \\
\hline\\
$^{90}$Zr   &4.9075   & 0.523   &  37.6  &35.5     & 6.13       & 0.623     &4.91   &[2.186, 0.061] &[2.740, 0.056]\\\
$^{92}$Zr   &4.9583   & 0.523   &  35.4  &38.8     & 6.02       & 0.687     &4.95   &[0.934, 0.083] &[2.339, 0.075]\\\
$^{92}$Mo   &4.9754   & 0.523   &  32.4  &40.4     & 6.04       & 0.610     &4.98   &[1.509, 0.097] &[2.849, 0.077]

\end{tabular}
\end{ruledtabular}
\end{table*}

Starting with the transition densities, the real term of the transition potential was obtained using the computer code DOLFIN \cite{DOLFIN}, whereas the imaginary term of the transition potential was obtained from analytical differentiation of the Woods-Saxon potential multiplied by the corresponding deformation length. DWBA cross sections for each excitation energy ($E_{x}$) were obtained for natural parities of multipolarities  from L=0 to 8. We used transition densities and sum rules for various multipolarities as described in Refs. \cite{Harakeh_book,Satchler1987,Harakeh1981}. To determine the uncertainties in $a_L(E_x)$, the python implementation EMCEE for the Markov-Chain Monte-Carlo (MCMC) algorithm of Goodman and Weare was employed \cite{foreman-mackey,goodman-weare}. The strength of this algorithm lies in its invariance to certain linear transformations, as discussed in Ref. \cite{foreman-mackey}. Provided that the algorithm runs until independent sampling is achieved, the resulting projections of the multidimensional posterior probability distribution onto the parameter axes are independent of the probability distributions for the other fit parameters. In short, this invariance renders the resulting probability distributions insensitive to covariances within the parameter space, thus allowing for a reliable and statistically meaningful extraction of parameter uncertainties.

From the resulting probability distributions for the parameters, the centroid of the peak was quoted as the central value with the 68\% confidence interval quoted as the $\pm 34\%$ bounds in the parameter range. The probability distributions are roughly normal, and hence the bounds of all reported uncertainties should be interpreted as the $1\sigma$ confidence intervals of quantities in question.

The strength distributions are obtained from the experimentally-determined EWSR fraction ($a_{L}$) using the relations provided in Ref. \cite{Harakeh_book}.
It should be noted that although we employed calculated DWBA cross sections up to L=8 in the MDA, the strengths could be reliably obtained only up to L=3 due to the limited angular range. Typical MDA fits for energy bins of 15, 20, 25, and 30 MeV are shown in Figs. \ref{MDA_90Zr} and \ref{MDA_92Mo} for $^{90}$Zr and $^{92}$Mo, respectively, along with the contributions from the L = 0, 1, 2, and 3 multipoles.  

\section{\label{sec:level4}Results and Discussion}
\begin{figure} 
\centering\includegraphics [trim= 0.11mm 0.5mm 0.1mm 0.1mm,
angle=360, clip, height=0.55\textheight]{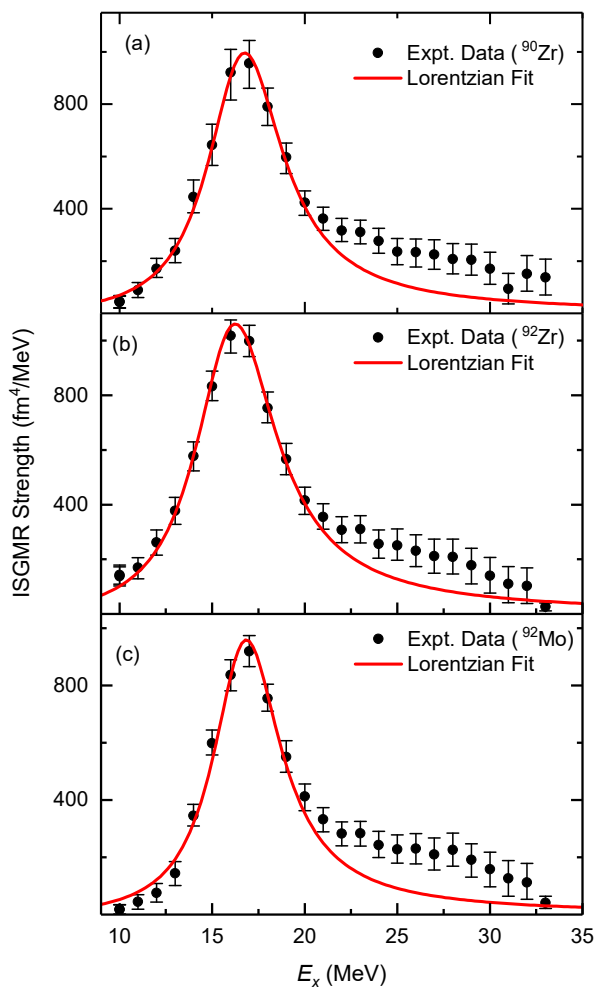}
\caption{(Color online) ISGMR strength distributions for $^{90}$Zr, $^{92}$Zr, and $^{92}$Mo (panels (a), (b), (c), respectively). The solid line in each panel is the Lorentzian fit to the data.}
\label{ISGMR_Indiv}
\end{figure}

We have extracted strength distributions for L=0, 1, and 2 over the excitation energy range of 9.5 to 32.5 MeV in all the three nuclei investigated in this work. ISGMR strength distributions, each consisting of a single and broad peak at around $E_{x}\sim$16.5 MeV, are displayed in Fig. \ref{ISGMR_Indiv}. The finite strength at higher excitation energies has been observed also previously in many nuclei, including in the Texas A \& M work  \cite{Li_2010, Uchida_90Zr, Darshana2012, YB_Mo2015, Krishi_2015}, and is attributable to the mimicking of the L=0 angular distribution by components of the nuclear continuum from the three-body channels, such as the forward-peaked knock-out process wherein protons and neutrons are knocked out by the incoming $\alpha$ projectiles \cite{BRANDENBURG1987}.
The ISGMR strength distributions are fitted with Lorentzian curves \cite{Berman_1975}:
\begin{equation}
\sigma (E)=\frac{\sigma_{m}}{1+\left(E^2-E^2_{m}\right)^2/E^2\Gamma^2}
\end{equation}

\noindent
where $E_{m}$ and $\Gamma$ are the peak energy and width of the resonance. The Lorentzian curve for each nucleus is shown by a solid line in the Fig. \ref{ISGMR_Indiv}. The Lorentzian parameters associated with the ISGMR ($E^\mathrm{L=0}_{m}$ and $\Gamma^\mathrm{L=0}$) for the three nuclei are very similar, as shown in Table \ref{ISGMR_Para}.

\begin{figure}[t] 
\centering\includegraphics [trim= 0.11mm 0.5mm 0.1mm 0.1mm,
angle=360, clip, height=0.52\textheight]{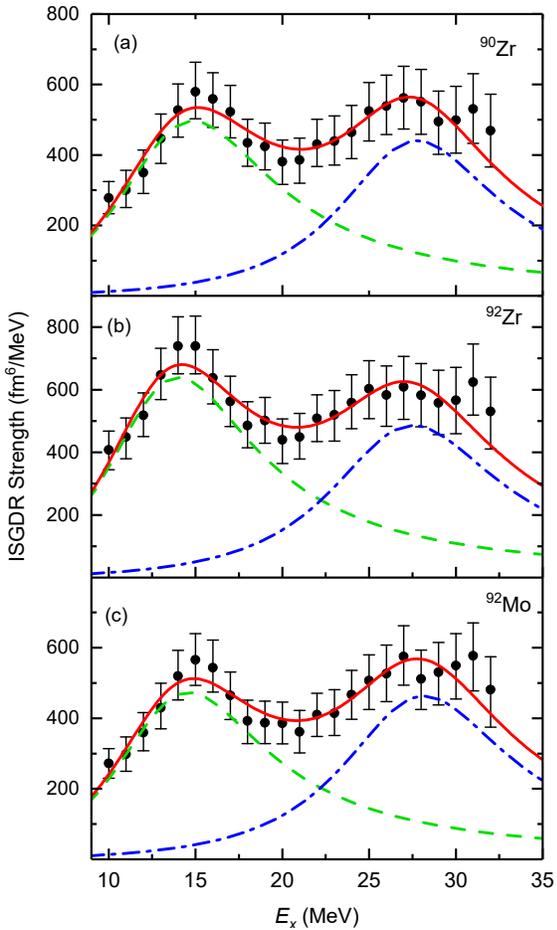}
\caption{(Color online) ISGDR strength distributions for $^{90}$Zr, $^{92}$Zr, and $^{92}$Mo (panels (a), (b), (c), respectively). The solid line in each panel is the two-peak fit to the data. The dashed and dash-dotted lines in each panel, respectively, represent the ``low-energy" and ``high-energy" components of the ``bimodal" ISGDR strength distribution. }
\label{ISGDR_Indiv}
\end{figure}

\begin{table*}
\caption{\label{ISGMR_Para} Lorentzian parameters and moment ratios for the ISGMR strength
distributions in $^{90,92}$Zr and $^{92}$Mo, where $m_{k}=\int E^{k}_{x} S(E_{x}) dE_{x}$ is the $k^{th}$ moment of the strength distribution. The results from TAMU (Refs. \cite{YB_Mo2015, Krishi_2015}) are provided for comparison.}
\begin{ruledtabular}
\begin{tabular}{ccccccccc}
  Nucleus     & $E^{L=0}_{m}$ & $\Gamma^{L=0}$  &EWSR & $m_{1}/m_{0}$  &$\sqrt{m_{1}/m_{-1}}$      &$\sqrt{m_{3}/m_{1}}$       &$K_{A}$ &Reference\\
              &      (MeV)    &          (MeV)  & (\%)& (MeV)          &                     (MeV) &                     (MeV) & (MeV)  &         \\
\hline\\
$^{90}$Zr     &  16.76 $\pm$ 0.12   &  4.96$^{+0.31}_{-0.32}$     & 74.7 $\pm$ 9   & 19.17$^{+0.21}_{-0.20}$  & 18.65 $^{+0.17}_{-0.17}$ & 20.87$^{+0.34}_{-0.33}$ &191.4 $\pm$ 6.1 & Present Work\footnotemark[1]\\
$^{90}$Zr     &  17.1               &  4.4     & 106 $\pm$ 12 & 17.88$^{+0.13}_{-0.11}$  & 17.58$^{+0.06}_{-0.04}$ & 18.86$^{+0.23}_{-0.14}$ &  & TAMU\footnotemark[2]\\
$^{92}$Zr     &  16.25 $\pm$ 0.10   &  5.33$^{+0.12}_{-0.20}$      & 78.9 $\pm$ 7 & 18.51$^{+0.17}_{-0.17}$  & 17.95$^{+0.15}_{-0.15}$ & 20.32$^{+0.27}_{-0.27}$ &184.5 $\pm$ 4.9& Present Work \footnotemark[1]\\
$^{92}$Zr     &  16.6               &  4.4     & 103 $\pm$ 12 & 18.23$^{+0.15}_{-0.13}$  & 17.71$^{+0.09}_{-0.07}$ & 20.09$^{+0.31}_{-0.22}$ &  & TAMU\footnotemark[2]\\
$^{92}$Mo     &  16.85 $\pm$ 0.10   &  4.44$^{+0.26}_{-0.25}$      &64.2 $\pm$ 6 & 19.49$^{+0.18}_{-0.17}$  & 19.00$^{+0.15}_{-0.15}$ & 21.09$^{+0.27}_{-0.26}$ &199.8 $\pm$ 5.0& Present Work \footnotemark[1]\\
$^{92}$Mo     &  16.8               &  4.0     & 107 $\pm$ 13 & 19.62$^{+0.28}_{-0.19}$  &                   & 21.68$^{+0.53}_{-0.33}$ &  & TAMU\footnotemark[2]\\
\end{tabular}
\footnotetext[1]{The moment ratios and EWSRs have been obtained over the $E_{x}$ ranges 10--30 MeV and 10--22 MeV (comprising the ISGMR peak), respectively.}
\footnotetext[2]{The TAMU work shows two-peak structure for the ISGMR strength distribution. Peak positions and widths (FWHM) correspond to the Gaussian fits of the low-energy peak. Moment ratios and EWSRs correspond to the $E_{x}$ range 10--35 MeV.}
\end{ruledtabular}
\end{table*}
\begin{figure}
\centering\includegraphics [trim= 0.11mm 0.5mm 0.1mm 0.1mm,angle=360, clip, width=0.37\textheight]{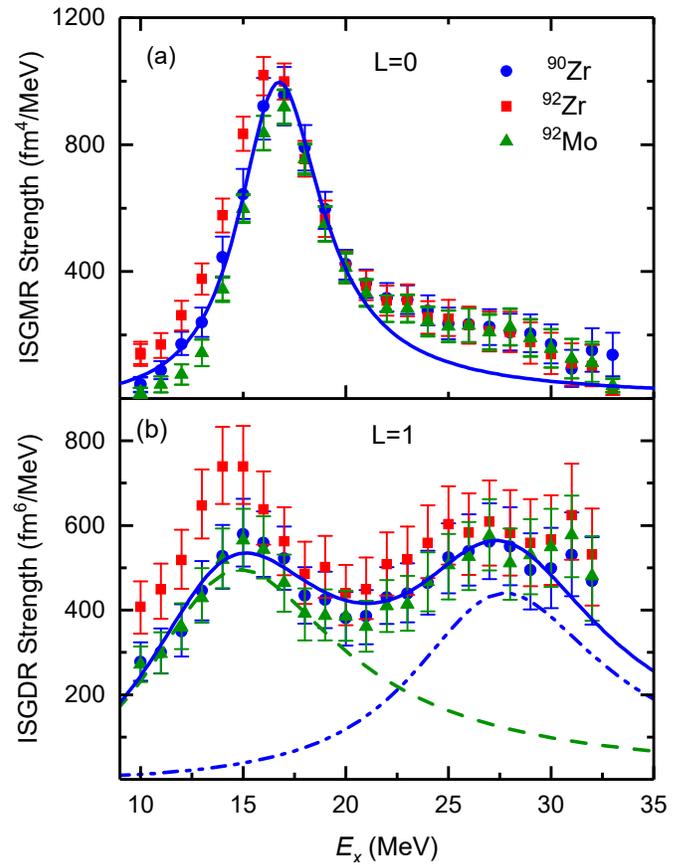}
\caption{(Color online) ISGMR (panel (a)) and ISGDR (panel (b)) strength distributions for  $^{90}$Zr (blue circles), $^{92}$Zr (red squares), and $^{92}$Mo (green triangles). The solid line in panel (a) represents the Lorentzian fit to the data for $^{90}$Zr. In panel (b), solid line  is the two-peak Lorentzian fit to the data, the dashed and dash-double-dotted  lines, respectively, represent the ``low-energy" and ``high-energy" components of the ``bimodal" ISGDR strength distribution for $^{90}$Zr (see text).}
\label{ISGMR_ISGDR}
\end{figure}

Three sum rules are generally used to quantify the giant-resonance strengths \cite{str1982}: (i) the polarizability sum rule ($m_{-1}$); (ii)  the energy-weighted sum rule (EWSR, $m_{1}$); and (iii)  the cubic energy-weighted sum rule ($m_{3}$), where $m_{k}=\int E^{k}_{x} S(E_{x}) dE_{x}$ and $S(E_{x})$ is the underlying strength distribution. The ratio of these different sum rules, $\sqrt{m_{3}/m_{1}}$ (in generalized scaling model) and $\sqrt{m_{1}/m_{-1}}$ (in hydrodynamical model), connects well with the centroid energy (or the collective frequency) of the compressional modes. Different moment ratios for the ISGMR strength distributions, calculated over the $E_{x}$ range 10--30 MeV are presented in Table \ref{ISGMR_Para}. The quoted EWSR fractions have been calculated over the excitation-energy range 10--22 MeV, encompassing the main ISGMR peak.
The quoted uncertainties in \%EWSR values are only statistical and do not include the systematic uncertainties (up to $\sim$20\%) arising from DWBA calculations, including those attributable to the choice of OM parameters (see, e.g., Ref. \cite{Li_2010}). Within the experimental uncertainties, different moment ratios for the three nuclei investigated in the present work are also nearly the same. The compressional modulus, $K_{A}$, determined within the scaling model ($\sqrt{m_{3}/m_{1}}$) using Eq. \ref{EISGMR} is also observed to be the same ($\sim$195 MeV) for all three nuclei (see Table \ref{ISGMR_Para}). It should be noted that the \%EWSR values shown here are a bit lower than those presented in Ref. \cite{YKGPLB2016}. This is a consequence of a more accurate accounting of the IVGDR contributions in the present analysis. The conclusions presented in that work remain unaffected in every manner, however. We also note that all three moment ratios for $^{92}$Zr and $^{92}$Mo deduced in the TAMU work, also included in Table \ref{ISGMR_Para},  overlap with those of present work, but the TAMU values for ISGMR in $^{90}$Zr are significantly lower. These differences are discussed later in the paper.

\begin{table}[!h]
\caption{\label{ISGDR_Para} Lorentzian parameters for the HE component of the ISGDR strength
distributions in $^{90,92}$Zr and $^{92}$Mo.  The results from TAMU (Refs. \cite{YB_Mo2015, Krishi_2015}) are provided for comparison. The EWSR fractions are calculated over the $E_{x}$ range 20--35 MeV. In the TAMU work, peak positions and widths (FWHM) correspond to Gaussian fits.}
\begin{ruledtabular}
\begin{tabular}{ccccc}
  Nucleus     & $E^{L=1}_{m}$ & $\Gamma^{L=1}$  &EWSR  &Reference\\
              &      (MeV)    &          (MeV)  & (\%) &         \\
\hline\\
$^{90}$Zr     &  27.76$^{+0.98}_{-0.78}$   &  11.28$^{+2.42}_{-2.70}$     & 68.7$^{+12.0}_{-12.0}$ &Present Work \\
$^{90}$Zr     &  27.4 $\pm$ 0.5     &  10.1 $\pm$ 2.0     & 49 $\pm$ 6 &TAMU  Work\\
$^{92}$Zr     &  27.53$^{+1.04}_{-0.86}$   &  12.09$^{+1.99}_{-2.59}$     & 74.3$^{+13.0}_{-13.0}$ &Present Work  \\
$^{92}$Zr     &  30.0 $\pm$ 0.7     &  12.9 $\pm$ 2.0     & 51 $\pm$ 7 &TAMU Work\\
$^{92}$Mo     &  28.16$^{+0.94}_{-0.82}$   &  11.92$^{+2.07}_{-2.57}$     & 65.4$^{+10.0}_{-11.0}$ &Present Work \\
$^{92}$Mo     &  27.6 $\pm$ 0.5     &  10.2 $\pm$ 2.0     & 59 $\pm$ 7 &TAMU Work\\
\end{tabular}
\end{ruledtabular}
\end{table}

ISGDR strength distributions, each consisting of two peaks, one at $\sim$15 MeV and another at $\sim$27 MeV, are shown in 
Fig.~\ref{ISGDR_Indiv} for all the three nuclei investigated in the present work. This bimodal pattern for the ISGDR has been observed in all nuclei investigated so far in both the RCNP and the TAMU measurements. The nature of the ``low-energy" (LE) peak at $\sim$15 MeV is not fully understood and has been suggested to correspond to ``toroidal'' \cite{kiev,Vretenar2000} or ``vortex'' modes \cite{Nesterenko2011, Nesterenko2014}. The centroid energy of this LE component of L=1 strength is independent of the nuclear incompressibility. The ``higher-energy" (HE) peak at $\sim$27 MeV  corresponds to the 3$\hbar\omega$ component of the L=1 compressional mode. The centroid energy of this HE component is directly related to the nuclear incompressibility, $K_{A}$, through Eq. (\ref{EISGDR}). The Lorentzian parameters of the HE component of the L=1 compressional mode, $E^{L=1}_{m}$ and $\Gamma^{L=1}$, are presented in Table \ref{ISGDR_Para}. One can see that, within the experimental uncertainties, both $E^{L=1}_{m}$ and $\Gamma^{L=1}$ are nearly the same for all the three nuclei. 
Also, except for $^{92}$Zr, the peak values for the HE components in the TAMU work, also shown in Table \ref{ISGDR_Para}, are not inconsistent with those in the present work. For $^{92}$Zr, however, the HE component of ISGDR in the TAMU work is $\sim$2 MeV higher than the other two nuclei; the reasons for this difference are not readily apparent.

\begin{table*}[t]
\caption{\label{ISGQR_Para} Lorentzian parameters, $m_{1}/m_{0}$, and EWSR fractions for the ISGQR strength
distributions in $^{90,92}$Zr and $^{92}$Mo.  The results from TAMU (Refs. \cite{YB_Mo2015, Krishi_2015}) are provided for comparison.}
\begin{ruledtabular}
\begin{tabular}{cccccc}
Nucleus      & $E^{L=2}_{m}$  (MeV)    & $\Gamma^{L=2}$  &$m_{1}/m_{0}$  (MeV)  &EWSR  & Reference\\
\hline\\
$^{90}$Zr     &  13.99 $\pm$ 0.07   &  7.44$^{+0.30}_{-0.28}$   &14.64$^{+0.22}_{-0.21}$   & 107.6 $\pm$ 5.0  & Present Work\footnotemark[1]\\
$^{90}$Zr     &  14.56 $\pm$ 0.20   &  4.94 $\pm$ 0.20   &14.09 $\pm$ 0.20   & 92 $\pm$ 12 & TAMU Work\footnotemark[2]\\
$^{92}$Zr     &  13.75 $\pm$ 0.07   &  7.59$^{+0.32}_{-0.29}$   &14.52$^{+0.18}_{-0.18}$   & 108.9 $\pm$ 5.0 & Present Work\footnotemark[1]\\
$^{92}$Zr     &  14.35 $\pm$ 0.15   &  4.8  $\pm$ 0.2    &14.16 $\pm$ 0.21  & 93 $\pm$ 12 & TAMU Work\footnotemark[2]\\
$^{92}$Mo     &  13.78 $\pm$ 0.07   &  7.75$^{+0.31}_{-0.28}$   &14.60$^{+0.18}_{-0.18}$   & 101.1 $\pm$ 5.0  & Present Work\footnotemark[1]\\
$^{92}$Mo     &  14.51 $\pm$ 0.23   &  4.84 $\pm$ 0.35   &14.16  $\pm$ 0.25  & 73 $\pm$ 13  & TAMU Work\footnotemark[1]\\
\end{tabular}
\footnotetext[1]{$m_{1}/m_{0}$ and EWSRs are calculated over the $E_{x}$ range 10--20 MeV (comprising the ISGQR peak).}
\footnotetext[2]{$m_{1}/m_{0}$ and EWSRs are calculated over the $E_{x}$ range 10--35 MeV. Peak positions and widths (FWHM) in the TAMU work  correspond to Gaussian fits.}
\end{ruledtabular}
\end{table*}
 
The ISGMR and ISGDR strength distributions for the three nuclei are overlaid in Figs. \ref{ISGMR_ISGDR}(a) and \ref{ISGMR_ISGDR}(b). One can see that the strength distributions of the three nuclei coincide with each other within experimental uncertainties for both the compressional modes, the ISGMR (Fig. \ref{ISGMR_ISGDR}(a)) and the HE component of the ISGDR (Fig. \ref{ISGMR_ISGDR}(b)). In the higher excitation energy region ($E_{x}$ = 20--30 MeV), where the results in Refs. \cite{YB_ZrMo2013, YB_Mo2015, Krishi_2015} had shown marked deviations for the three nuclei, the strength distributions for $^{92}$Zr and $^{92}$Mo observed in the present work are identical to that in $^{90}$Zr, not only for the ISGMR but also for the ISGDR. The LE component of the ISGDR, however, shows marked differences for the three nuclei; in particular, $^{92}$Zr shows more LE strength than the remaining two. This might provide input to the theoretical studies of ``toroidal'' \cite{kiev,Vretenar2000} or  ``vortex'' modes \cite{Nesterenko2011, Nesterenko2014}, which have not been conclusively established in experiments yet. 

\begin{figure}[!h]
\centering\includegraphics [trim= 0.11mm 0.5mm 0.1mm 0.1mm,
angle=360, clip, height=0.55\textheight]{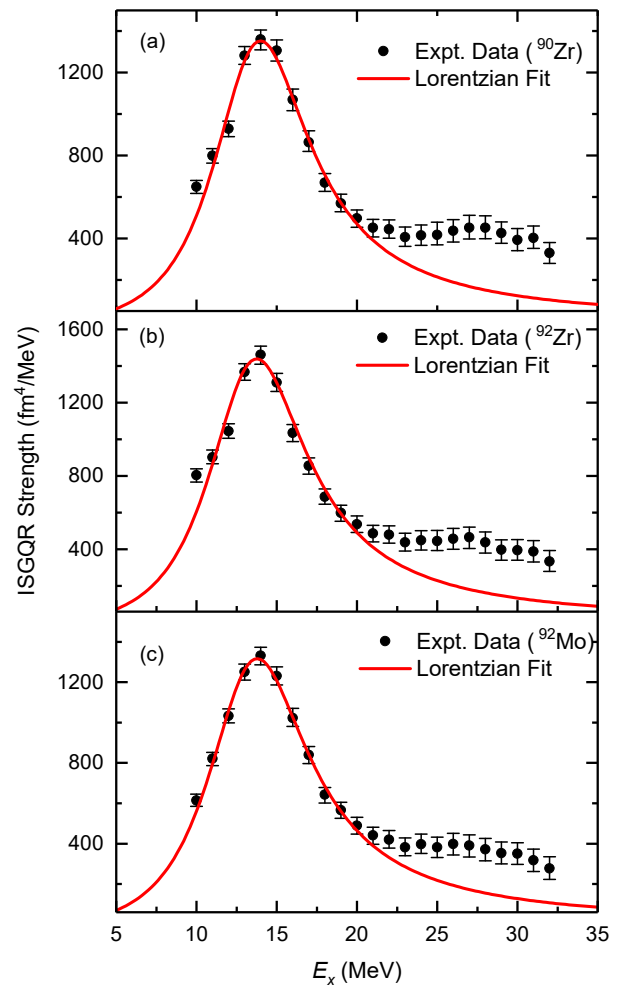}
\caption{(Color online) ISGQR strength distributions for $^{90}$Zr, $^{92}$Zr, and $^{92}$Mo (panels (a), (b), (c), respectively). The solid line in each panel is the Lorentzian fit to the data.}
\label{ISGQR_Indiv}
\end{figure}

The ISGQR strength distributions for the three nuclei, each consisting of a broad peak at $\sim$13.5 MeV, are shown in the Fig. \ref{ISGQR_Indiv}, where the solid line in each panel represents the Lorentzian fit to the data.  The fit parameters, $E^{L=2}_{m}$ and $\Gamma^{L=2}$ are presented in Table \ref{ISGQR_Para}. The EWSR fraction and moment ratio $m_{1}/m_{0}$ for each nucleus are determined in the $E_{x}$ range 10--20 MeV, comprising the full ISGQR peak. The moment ratio $m_{1}/m_{0}$ for the three nuclei coincide within the experimental uncertainties. The EWSR fraction for the ISGQR is obtained to be $\sim$100\% for all the three nuclei. So, the ISGQR strengths, although not related to nuclear compressibility, coincide as well for the three nuclei within the experimental uncertainties. For this mode, the TAMU results are very similar to those obtained in this work.

Thus, we observe qualitatively, from the zero degree and difference spectra, as well as quantitatively, from ISGMR, ISGDR, and ISGQR strength distributions, that nuclei investigated in the present work behave in a nearly identical manner, as far as the collective excitations are concerned. In fact, there had been no report prior to the results presented in Refs. \cite{YB_ZrMo2013, YB_Mo2015, Krishi_2015} of any ``shell effects" leading to significant differences between ISGMR energies in nearby nuclei. For instance, measurements on three Lead isotopes,  $^{204,206,208}$Pb, had resulted in very similar ISGMR energies \cite{Darshna2013}. Further, detailed investigations of the ISGMR over the Sn and Cd isotopic chains have been performed in recent years \cite{Li_PRL2007, Li_2010, Darshana2012}. Although these nuclei emerged as ``soft" in comparison to the ``standard" nuclei, $^{90}$Zr and $^{208}$Pb, the ISGMR energies varied quite smoothly over a wide range of the asymmetry parameter $(N-Z)/A$ \cite{Li_2010}.

The obvious question is why the present results are so different from those obtained by the Texas A \& M group \cite{YB_ZrMo2013, YB_Mo2015, Krishi_2015}. As stated previously in Ref. \cite{YKGPLB2016}, we believe the answer lies in the way the background is accounted for in the two approaches. In the present work, all instrumental background is eliminated because of the superior optical properties of the Grand Raiden Spectrometer (see, e.g., Ref. \cite{Itoh_prc2003}, and Fig. \ref{ZeroDeg}), leaving only the physical continuum as part of the excitation-energy spectra. In the Texas A \& M work, an empirical background is subtracted by assuming that it has the shape of a straight line at high excitation, joining onto a Fermi shape at low excitation to model particle threshold effects \cite{YB_116Sn_2004,dhybg2}. This process subtracts the physical continuum as well. It is quite possible, and perhaps likely, that this background subtraction approach is responsible for the different ISGMR strengths observed for various nuclei in their work. Since there is no arbitrariness involved in the background-subtraction procedure employed in the present work, it may be argued that our final results are more reliable.

\section{\label{sec:level5}Summary}
We have investigated the ISGMR, ISGDR, and ISGQR response in $^{90, 92}$Zr and $^{92}$Mo via inelastic scattering of 385-MeV $\alpha$ particles at extremely forward angles (including 0$^\circ$). In contrast with recent reports where significant differences were observed in the ISGMR strength distributions for $^{92}$Zr and $^{92}$Mo as compared with that for $^{90}$Zr, not only the  ISGMR response of these nuclei but also the ISGDR and ISGQR responses, are found to be practically identical. These results affirm the standard hydrodynamical picture associated with collective modes of oscillation, and clearly indicate that the compression modes ISGMR and ISGDR and, hence, the nuclear incompressibility, are not influenced by the shell structure of the nuclei near $A\sim$90. 

\section{\label{sec:level6}Acknowledgments}
The authors acknowledge the efforts of the staff of the RCNP Ring Cyclotron Facility in providing a high-quality, halo-free $\alpha$ beam required for the measurements reported here. This work has been supported in part by the U.S. National Science Foundation (Grants No. PHY-1419765 and No. PHY-1713857). M.~{\c S}enyi{\u g}it was supported by the Scientific
and Technical Council of Turkey (T\"{U}B\.{I}TAK) under the  B\.{I}DEB 2219 Scholarship program.

\begin{thebibliography}{52}
\expandafter\ifx\csname natexlab\endcsname\relax\def\natexlab#1{#1}\fi
\expandafter\ifx\csname bibnamefont\endcsname\relax
  \def\bibnamefont#1{#1}\fi
\expandafter\ifx\csname bibfnamefont\endcsname\relax
  \def\bibfnamefont#1{#1}\fi
\expandafter\ifx\csname citenamefont\endcsname\relax
  \def\citenamefont#1{#1}\fi
\expandafter\ifx\csname url\endcsname\relax
  \def\url#1{\texttt{#1}}\fi
\expandafter\ifx\csname urlprefix\endcsname\relax\def\urlprefix{URL }\fi
\providecommand{\bibinfo}[2]{#2}
\providecommand{\eprint}[2][]{\url{#2}}

\bibitem[{\citenamefont{Harakeh and van~der Woude}(2001)}]{Harakeh_book}
\bibinfo{author}{\bibfnamefont{M.~N.} \bibnamefont{Harakeh}} \bibnamefont{and}
  \bibinfo{author}{\bibfnamefont{A.}~\bibnamefont{van~der Woude}},
  \emph{\bibinfo{title}{Giant Resonances Fundamental High-Frequency Modes of
  Nuclear Excitation}} (\bibinfo{publisher}{Oxford University Press, New York},
  \bibinfo{year}{2001}).

\bibitem[{\citenamefont{Lattimer and Prakash}(2001)}]{Lattimer2001}
\bibinfo{author}{\bibfnamefont{J.~M.} \bibnamefont{Lattimer}} \bibnamefont{and}
  \bibinfo{author}{\bibfnamefont{M.}~\bibnamefont{Prakash}},
  \bibinfo{journal}{Astrophys. J.} \textbf{\bibinfo{volume}{550}},
  \bibinfo{pages}{426} (\bibinfo{year}{2001}).

\bibitem[{\citenamefont{Stringari}(1982)}]{str1982}
\bibinfo{author}{\bibfnamefont{S.}~\bibnamefont{Stringari}},
  \bibinfo{journal}{Phys. Lett. B} \textbf{\bibinfo{volume}{108}},
  \bibinfo{pages}{232} (\bibinfo{year}{1982}).

\bibitem[{\citenamefont{Treiner et~al.}(1981)\citenamefont{Treiner, Krivine,
  Bohigas, and Martorell}}]{Treiner1981}
\bibinfo{author}{\bibfnamefont{J.}~\bibnamefont{Treiner}},
  \bibinfo{author}{\bibfnamefont{H.}~\bibnamefont{Krivine}},
  \bibinfo{author}{\bibfnamefont{O.}~\bibnamefont{Bohigas}}, \bibnamefont{and}
  \bibinfo{author}{\bibfnamefont{J.}~\bibnamefont{Martorell}},
  \bibinfo{journal}{Nucl. Phys. A} \textbf{\bibinfo{volume}{371}},
  \bibinfo{pages}{253} (\bibinfo{year}{1981}).

\bibitem[{\citenamefont{Blaizot et~al.}(1995)\citenamefont{Blaizot, Berger,
  Decharg\'e, and Girod}}]{Blaizot1995}
\bibinfo{author}{\bibfnamefont{J.~P.} \bibnamefont{Blaizot}},
  \bibinfo{author}{\bibfnamefont{J.~F.} \bibnamefont{Berger}},
  \bibinfo{author}{\bibfnamefont{J.}~\bibnamefont{Decharg\'e}},
  \bibnamefont{and} \bibinfo{author}{\bibfnamefont{M.}~\bibnamefont{Girod}},
  \bibinfo{journal}{Nucl. Phys. A} \textbf{\bibinfo{volume}{591}},
  \bibinfo{pages}{435} (\bibinfo{year}{1995}).

\bibitem[{\citenamefont{Col\`o et~al.}(2004)\citenamefont{Col\`o, Giai, Meyer,
  Bennaceur, and Bonche}}]{Colo_Giai_PRC2004}
\bibinfo{author}{\bibfnamefont{G.}~\bibnamefont{Col\`o}},
  \bibinfo{author}{\bibfnamefont{N.~V.} \bibnamefont{Giai}},
  \bibinfo{author}{\bibfnamefont{J.}~\bibnamefont{Meyer}},
  \bibinfo{author}{\bibfnamefont{K.}~\bibnamefont{Bennaceur}},
  \bibnamefont{and} \bibinfo{author}{\bibfnamefont{P.}~\bibnamefont{Bonche}},
  \bibinfo{journal}{Phys. Rev. C} \textbf{\bibinfo{volume}{70}},
  \bibinfo{pages}{024307} (\bibinfo{year}{2004}).

\bibitem[{\citenamefont{Todd-Rutel and Piekarewicz}(2005)}]{Todd-Rutel_PRL2005}
\bibinfo{author}{\bibfnamefont{B.~G.} \bibnamefont{Todd-Rutel}}
  \bibnamefont{and}
  \bibinfo{author}{\bibfnamefont{J.}~\bibnamefont{Piekarewicz}},
  \bibinfo{journal}{Phys. Rev. Lett.} \textbf{\bibinfo{volume}{95}},
  \bibinfo{pages}{122501} (\bibinfo{year}{2005}).

\bibitem[{\citenamefont{Col\`o et~al.}(2014)\citenamefont{Col\`o, Garg, and
  Sagawa}}]{colo_garg_sagawa}
\bibinfo{author}{\bibfnamefont{G.}~\bibnamefont{Col\`o}},
  \bibinfo{author}{\bibfnamefont{U.}~\bibnamefont{Garg}}, \bibnamefont{and}
  \bibinfo{author}{\bibfnamefont{H.}~\bibnamefont{Sagawa}},
  \bibinfo{journal}{Eur. Phys. J. A} \textbf{\bibinfo{volume}{50}},
  \bibinfo{pages}{26} (\bibinfo{year}{2014}).

\bibitem[{\citenamefont{Piekarewicz}(2014)}]{pickarewickz2014}
\bibinfo{author}{\bibfnamefont{J.}~\bibnamefont{Piekarewicz}},
  \bibinfo{journal}{Eur. Phys. J. A} \textbf{\bibinfo{volume}{50}},
  \bibinfo{pages}{25} (\bibinfo{year}{2014}).

\bibitem[{\citenamefont{Uchida et~al.}(2004)\citenamefont{Uchida, Sakaguchi,
  Itoh, Yosoi, Kawabata, Yasuda, Takeda, Murakami, Terashima, Kishi
  et~al.}}]{Uchida_90Zr}
\bibinfo{author}{\bibfnamefont{M.}~\bibnamefont{Uchida}},
  \bibinfo{author}{\bibfnamefont{H.}~\bibnamefont{Sakaguchi}},
  \bibinfo{author}{\bibfnamefont{M.}~\bibnamefont{Itoh}},
  \bibinfo{author}{\bibfnamefont{M.}~\bibnamefont{Yosoi}},
  \bibinfo{author}{\bibfnamefont{T.}~\bibnamefont{Kawabata}},
  \bibinfo{author}{\bibfnamefont{Y.}~\bibnamefont{Yasuda}},
  \bibinfo{author}{\bibfnamefont{H.}~\bibnamefont{Takeda}},
  \bibinfo{author}{\bibfnamefont{T.}~\bibnamefont{Murakami}},
  \bibinfo{author}{\bibfnamefont{S.}~\bibnamefont{Terashima}},
  \bibinfo{author}{\bibfnamefont{S.}~\bibnamefont{Kishi}},
  \bibnamefont{et~al.}, \bibinfo{journal}{Phys. Rev. C}
  \textbf{\bibinfo{volume}{69}}, \bibinfo{pages}{051301}
  (\bibinfo{year}{2004}).

\bibitem[{\citenamefont{Uchida et~al.}(2003)\citenamefont{Uchida, Sakaguchi,
  Itoh, Yosoi, Kawabata, Takeda, Yasuda, Murakami, Ishikawa, Taki
  et~al.}}]{Uchida_PLB2003}
\bibinfo{author}{\bibfnamefont{M.}~\bibnamefont{Uchida}},
  \bibinfo{author}{\bibfnamefont{H.}~\bibnamefont{Sakaguchi}},
  \bibinfo{author}{\bibfnamefont{M.}~\bibnamefont{Itoh}},
  \bibinfo{author}{\bibfnamefont{M.}~\bibnamefont{Yosoi}},
  \bibinfo{author}{\bibfnamefont{T.}~\bibnamefont{Kawabata}},
  \bibinfo{author}{\bibfnamefont{H.}~\bibnamefont{Takeda}},
  \bibinfo{author}{\bibfnamefont{Y.}~\bibnamefont{Yasuda}},
  \bibinfo{author}{\bibfnamefont{T.}~\bibnamefont{Murakami}},
  \bibinfo{author}{\bibfnamefont{T.}~\bibnamefont{Ishikawa}},
  \bibinfo{author}{\bibfnamefont{T.}~\bibnamefont{Taki}}, \bibnamefont{et~al.},
  \bibinfo{journal}{Phys. Lett. B} \textbf{\bibinfo{volume}{557}},
  \bibinfo{pages}{12} (\bibinfo{year}{2003}).

\bibitem[{\citenamefont{Itoh et~al.}(2003)\citenamefont{Itoh, Sakaguchi,
  Uchida, Ishikawa, Kawabata, Murakami, Takeda, Taki, Terashima, Tsukahara
  et~al.}}]{Itoh_prc2003}
\bibinfo{author}{\bibfnamefont{M.}~\bibnamefont{Itoh}},
  \bibinfo{author}{\bibfnamefont{H.}~\bibnamefont{Sakaguchi}},
  \bibinfo{author}{\bibfnamefont{M.}~\bibnamefont{Uchida}},
  \bibinfo{author}{\bibfnamefont{T.}~\bibnamefont{Ishikawa}},
  \bibinfo{author}{\bibfnamefont{T.}~\bibnamefont{Kawabata}},
  \bibinfo{author}{\bibfnamefont{T.}~\bibnamefont{Murakami}},
  \bibinfo{author}{\bibfnamefont{H.}~\bibnamefont{Takeda}},
  \bibinfo{author}{\bibfnamefont{T.}~\bibnamefont{Taki}},
  \bibinfo{author}{\bibfnamefont{S.}~\bibnamefont{Terashima}},
  \bibinfo{author}{\bibfnamefont{N.}~\bibnamefont{Tsukahara}},
  \bibnamefont{et~al.}, \bibinfo{journal}{Phys. Rev. C}
  \textbf{\bibinfo{volume}{68}}, \bibinfo{pages}{064602}
  (\bibinfo{year}{2003}).

\bibitem[{\citenamefont{Nayak et~al.}(2006)\citenamefont{Nayak, Garg, Hedden,
  Koss, Li, Liu, \mbox{P. V. Madhusudhana Rao}, Zhu, Itoh, Sakaguchi
  et~al.}}]{nayak2006}
\bibinfo{author}{\bibfnamefont{B.~K.} \bibnamefont{Nayak}},
  \bibinfo{author}{\bibfnamefont{U.}~\bibnamefont{Garg}},
  \bibinfo{author}{\bibfnamefont{M.}~\bibnamefont{Hedden}},
  \bibinfo{author}{\bibfnamefont{M.}~\bibnamefont{Koss}},
  \bibinfo{author}{\bibfnamefont{T.}~\bibnamefont{Li}},
  \bibinfo{author}{\bibfnamefont{Y.}~\bibnamefont{Liu}},
  \bibinfo{author}{\bibnamefont{\mbox{P. V. Madhusudhana Rao}}},
  \bibinfo{author}{\bibfnamefont{S.}~\bibnamefont{Zhu}},
  \bibinfo{author}{\bibfnamefont{M.}~\bibnamefont{Itoh}},
  \bibinfo{author}{\bibfnamefont{H.}~\bibnamefont{Sakaguchi}},
  \bibnamefont{et~al.}, \bibinfo{journal}{Phys. Lett. B}
  \textbf{\bibinfo{volume}{637}}, \bibinfo{pages}{43} (\bibinfo{year}{2006}).

\bibitem[{\citenamefont{Garg et~al.}(2007)\citenamefont{Garg, Li, Okumura,
  Akimune, Fujiwara, Harakeh, Hashimoto, Itoh, Iwao, Kawabata
  et~al.}}]{fluffy1}
\bibinfo{author}{\bibfnamefont{U.}~\bibnamefont{Garg}},
  \bibinfo{author}{\bibfnamefont{T.}~\bibnamefont{Li}},
  \bibinfo{author}{\bibfnamefont{S.}~\bibnamefont{Okumura}},
  \bibinfo{author}{\bibfnamefont{H.}~\bibnamefont{Akimune}},
  \bibinfo{author}{\bibfnamefont{M.}~\bibnamefont{Fujiwara}},
  \bibinfo{author}{\bibfnamefont{M.}~\bibnamefont{Harakeh}},
  \bibinfo{author}{\bibfnamefont{H.}~\bibnamefont{Hashimoto}},
  \bibinfo{author}{\bibfnamefont{M.}~\bibnamefont{Itoh}},
  \bibinfo{author}{\bibfnamefont{Y.}~\bibnamefont{Iwao}},
  \bibinfo{author}{\bibfnamefont{T.}~\bibnamefont{Kawabata}},
  \bibnamefont{et~al.}, \bibinfo{journal}{Nucl. Phys. A}
  \textbf{\bibinfo{volume}{788}}, \bibinfo{pages}{36} (\bibinfo{year}{2007}).

\bibitem[{\citenamefont{Li et~al.}(2007)\citenamefont{Li, Garg, Liu, Marks,
  Nayak, \mbox{Madhusudhana} Rao, Fujiwara, Hashimoto, Kawase, Nakanishi
  et~al.}}]{Li_PRL2007}
\bibinfo{author}{\bibfnamefont{T.}~\bibnamefont{Li}},
  \bibinfo{author}{\bibfnamefont{U.}~\bibnamefont{Garg}},
  \bibinfo{author}{\bibfnamefont{Y.}~\bibnamefont{Liu}},
  \bibinfo{author}{\bibfnamefont{R.}~\bibnamefont{Marks}},
  \bibinfo{author}{\bibfnamefont{B.~K.} \bibnamefont{Nayak}},
  \bibinfo{author}{\bibfnamefont{P.~V.} \bibnamefont{\mbox{Madhusudhana} Rao}},
  \bibinfo{author}{\bibfnamefont{M.}~\bibnamefont{Fujiwara}},
  \bibinfo{author}{\bibfnamefont{H.}~\bibnamefont{Hashimoto}},
  \bibinfo{author}{\bibfnamefont{K.}~\bibnamefont{Kawase}},
  \bibinfo{author}{\bibfnamefont{K.}~\bibnamefont{Nakanishi}},
  \bibnamefont{et~al.}, \bibinfo{journal}{Phys. Rev. Lett.}
  \textbf{\bibinfo{volume}{99}}, \bibinfo{pages}{162503}
  (\bibinfo{year}{2007}).

\bibitem[{\citenamefont{Li et~al.}(2010)\citenamefont{Li, Garg, Liu, Marks,
  Nayak, \mbox{Madhusudhana} Rao, Fujiwara, Hashimoto, Nakanishi, Okumura
  et~al.}}]{Li_2010}
\bibinfo{author}{\bibfnamefont{T.}~\bibnamefont{Li}},
  \bibinfo{author}{\bibfnamefont{U.}~\bibnamefont{Garg}},
  \bibinfo{author}{\bibfnamefont{Y.}~\bibnamefont{Liu}},
  \bibinfo{author}{\bibfnamefont{R.}~\bibnamefont{Marks}},
  \bibinfo{author}{\bibfnamefont{B.~K.} \bibnamefont{Nayak}},
  \bibinfo{author}{\bibfnamefont{P.~V.} \bibnamefont{\mbox{Madhusudhana} Rao}},
  \bibinfo{author}{\bibfnamefont{M.}~\bibnamefont{Fujiwara}},
  \bibinfo{author}{\bibfnamefont{H.}~\bibnamefont{Hashimoto}},
  \bibinfo{author}{\bibfnamefont{K.}~\bibnamefont{Nakanishi}},
  \bibinfo{author}{\bibfnamefont{S.}~\bibnamefont{Okumura}},
  \bibnamefont{et~al.}, \bibinfo{journal}{Phys. Rev. C}
  \textbf{\bibinfo{volume}{81}}, \bibinfo{pages}{034309}
  (\bibinfo{year}{2010}).

\bibitem[{\citenamefont{Patel et~al.}(2012)\citenamefont{Patel, Garg, Fujiwara,
  Akimune, Berg, Harakeh, Itoh, Kawabata, Kawase, Nayak et~al.}}]{Darshana2012}
\bibinfo{author}{\bibfnamefont{D.}~\bibnamefont{Patel}},
  \bibinfo{author}{\bibfnamefont{U.}~\bibnamefont{Garg}},
  \bibinfo{author}{\bibfnamefont{M.}~\bibnamefont{Fujiwara}},
  \bibinfo{author}{\bibfnamefont{H.}~\bibnamefont{Akimune}},
  \bibinfo{author}{\bibfnamefont{G.~P.~A.} \bibnamefont{Berg}},
  \bibinfo{author}{\bibfnamefont{M.~N.} \bibnamefont{Harakeh}},
  \bibinfo{author}{\bibfnamefont{M.}~\bibnamefont{Itoh}},
  \bibinfo{author}{\bibfnamefont{T.}~\bibnamefont{Kawabata}},
  \bibinfo{author}{\bibfnamefont{K.}~\bibnamefont{Kawase}},
  \bibinfo{author}{\bibfnamefont{B.~K.} \bibnamefont{Nayak}},
  \bibnamefont{et~al.}, \bibinfo{journal}{Phys. Lett. B}
  \textbf{\bibinfo{volume}{718}}, \bibinfo{pages}{447} (\bibinfo{year}{2012}).

\bibitem[{\citenamefont{Piekarewicz}(2007)}]{fluffy2}
\bibinfo{author}{\bibfnamefont{J.}~\bibnamefont{Piekarewicz}},
  \bibinfo{journal}{Phys. Rev. C} \textbf{\bibinfo{volume}{76}},
  \bibinfo{pages}{031301} (\bibinfo{year}{2007}).

\bibitem[{\citenamefont{Tselyaev et~al.}(2009)\citenamefont{Tselyaev, Speth,
  Krewald, Litvinova, Kamerdzhiev, Lyutorovich, Avdeenkov, and
  Gr\"{u}mmer}}]{Speth2009}
\bibinfo{author}{\bibfnamefont{V.}~\bibnamefont{Tselyaev}},
  \bibinfo{author}{\bibfnamefont{J.}~\bibnamefont{Speth}},
  \bibinfo{author}{\bibfnamefont{S.}~\bibnamefont{Krewald}},
  \bibinfo{author}{\bibfnamefont{E.}~\bibnamefont{Litvinova}},
  \bibinfo{author}{\bibfnamefont{S.}~\bibnamefont{Kamerdzhiev}},
  \bibinfo{author}{\bibfnamefont{N.}~\bibnamefont{Lyutorovich}},
  \bibinfo{author}{\bibfnamefont{A.}~\bibnamefont{Avdeenkov}},
  \bibnamefont{and}
  \bibinfo{author}{\bibfnamefont{F.}~\bibnamefont{Gr\"{u}mmer}},
  \bibinfo{journal}{Phys. Rev. C} \textbf{\bibinfo{volume}{79}},
  \bibinfo{pages}{034309} (\bibinfo{year}{2009}).

\bibitem[{\citenamefont{Piekarewicz}(2010)}]{fluffy6}
\bibinfo{author}{\bibfnamefont{J.}~\bibnamefont{Piekarewicz}},
  \bibinfo{journal}{J. Phys. G: Nucl. Part. Phys.}
  \textbf{\bibinfo{volume}{37}}, \bibinfo{pages}{064038}
  (\bibinfo{year}{2010}).

\bibitem[{\citenamefont{\mbox{Li-Gang Cao}
  et~al.}(2012)\citenamefont{\mbox{Li-Gang Cao}, Sagawa, and Col\`o}}]{fluffy5}
\bibinfo{author}{\bibnamefont{\mbox{Li-Gang Cao}}},
  \bibinfo{author}{\bibfnamefont{H.}~\bibnamefont{Sagawa}}, \bibnamefont{and}
  \bibinfo{author}{\bibfnamefont{G.}~\bibnamefont{Col\`o}},
  \bibinfo{journal}{Phys. Rev. C} \textbf{\bibinfo{volume}{86}},
  \bibinfo{pages}{054313} (\bibinfo{year}{2012}).

\bibitem[{\citenamefont{Vesel\'y et~al.}(2012)\citenamefont{Vesel\'y, Toivanen,
  Carlsson, Dobaczewski, Michel, and Pastore}}]{fluffy4}
\bibinfo{author}{\bibfnamefont{P.}~\bibnamefont{Vesel\'y}},
  \bibinfo{author}{\bibfnamefont{J.}~\bibnamefont{Toivanen}},
  \bibinfo{author}{\bibfnamefont{B.~G.} \bibnamefont{Carlsson}},
  \bibinfo{author}{\bibfnamefont{J.}~\bibnamefont{Dobaczewski}},
  \bibinfo{author}{\bibfnamefont{N.}~\bibnamefont{Michel}}, \bibnamefont{and}
  \bibinfo{author}{\bibfnamefont{A.}~\bibnamefont{Pastore}},
  \bibinfo{journal}{Phys. Rev. C} \textbf{\bibinfo{volume}{86}},
  \bibinfo{pages}{024303} (\bibinfo{year}{2012}).

\bibitem[{\citenamefont{Youngblood et~al.}(2013)\citenamefont{Youngblood, Lui,
  Krishichayan, Button, Anders, Gorelik, Urin, and Shlomo}}]{YB_ZrMo2013}
\bibinfo{author}{\bibfnamefont{D.~H.} \bibnamefont{Youngblood}},
  \bibinfo{author}{\bibfnamefont{Y.-W.} \bibnamefont{Lui}},
  \bibinfo{author}{\bibnamefont{Krishichayan}},
  \bibinfo{author}{\bibfnamefont{J.}~\bibnamefont{Button}},
  \bibinfo{author}{\bibfnamefont{M.~R.} \bibnamefont{Anders}},
  \bibinfo{author}{\bibfnamefont{M.~L.} \bibnamefont{Gorelik}},
  \bibinfo{author}{\bibfnamefont{M.~H.} \bibnamefont{Urin}}, \bibnamefont{and}
  \bibinfo{author}{\bibfnamefont{S.}~\bibnamefont{Shlomo}},
  \bibinfo{journal}{Phys. Rev. C} \textbf{\bibinfo{volume}{88}},
  \bibinfo{pages}{021301} (\bibinfo{year}{2013}).

\bibitem[{\citenamefont{Youngblood et~al.}(2015)\citenamefont{Youngblood, Lui,
  Krishichayan, Button, Bonasera, and Shlomo}}]{YB_Mo2015}
\bibinfo{author}{\bibfnamefont{D.~H.} \bibnamefont{Youngblood}},
  \bibinfo{author}{\bibfnamefont{Y.-W.} \bibnamefont{Lui}},
  \bibinfo{author}{\bibnamefont{Krishichayan}},
  \bibinfo{author}{\bibfnamefont{J.}~\bibnamefont{Button}},
  \bibinfo{author}{\bibfnamefont{G.}~\bibnamefont{Bonasera}}, \bibnamefont{and}
  \bibinfo{author}{\bibfnamefont{S.}~\bibnamefont{Shlomo}},
  \bibinfo{journal}{Phys. Rev. C} \textbf{\bibinfo{volume}{92}},
  \bibinfo{pages}{014318} (\bibinfo{year}{2015}).

\bibitem[{\citenamefont{Krishichayan et~al.}(2015)\citenamefont{Krishichayan,
  Lui, Button, Youngblood, Bonasera, and Shlomo}}]{Krishi_2015}
\bibinfo{author}{\bibnamefont{Krishichayan}},
  \bibinfo{author}{\bibfnamefont{Y.-W.} \bibnamefont{Lui}},
  \bibinfo{author}{\bibfnamefont{J.}~\bibnamefont{Button}},
  \bibinfo{author}{\bibfnamefont{D.~H.} \bibnamefont{Youngblood}},
  \bibinfo{author}{\bibfnamefont{G.}~\bibnamefont{Bonasera}}, \bibnamefont{and}
  \bibinfo{author}{\bibfnamefont{S.}~\bibnamefont{Shlomo}},
  \bibinfo{journal}{Phys. Rev. C} \textbf{\bibinfo{volume}{92}},
  \bibinfo{pages}{044323} (\bibinfo{year}{2015}).

\bibitem[{\citenamefont{Button et~al.}(2016)\citenamefont{Button, Lui,
  Youngblood, Chen, Bonasera, and Shlomo}}]{Button2016}
\bibinfo{author}{\bibfnamefont{J.}~\bibnamefont{Button}},
  \bibinfo{author}{\bibfnamefont{Y.-W.} \bibnamefont{Lui}},
  \bibinfo{author}{\bibfnamefont{D.~H.} \bibnamefont{Youngblood}},
  \bibinfo{author}{\bibfnamefont{X.}~\bibnamefont{Chen}},
  \bibinfo{author}{\bibfnamefont{G.}~\bibnamefont{Bonasera}}, \bibnamefont{and}
  \bibinfo{author}{\bibfnamefont{S.}~\bibnamefont{Shlomo}},
  \bibinfo{journal}{Phys. Rev. C} \textbf{\bibinfo{volume}{94}},
  \bibinfo{pages}{034315} (\bibinfo{year}{2016}).

\bibitem[{\citenamefont{Harakeh et~al.}(1977)\citenamefont{Harakeh, van~der
  Borg, Ishimatsu, Morsch, van~der Woude, and Bertrand}}]{Harakesh_prl1977}
\bibinfo{author}{\bibfnamefont{M.~N.} \bibnamefont{Harakeh}},
  \bibinfo{author}{\bibfnamefont{K.}~\bibnamefont{van~der Borg}},
  \bibinfo{author}{\bibfnamefont{T.}~\bibnamefont{Ishimatsu}},
  \bibinfo{author}{\bibfnamefont{H.~P.} \bibnamefont{Morsch}},
  \bibinfo{author}{\bibfnamefont{A.}~\bibnamefont{van~der Woude}},
  \bibnamefont{and} \bibinfo{author}{\bibfnamefont{F.~E.}
  \bibnamefont{Bertrand}}, \bibinfo{journal}{Phys. Rev. Lett}
  \textbf{\bibinfo{volume}{38}}, \bibinfo{pages}{676} (\bibinfo{year}{1977}).

\bibitem[{\citenamefont{Youngblood et~al.}(1977)\citenamefont{Youngblood,
  Rozsa, Moss, Brown, and Bronson}}]{YBPRL1977}
\bibinfo{author}{\bibfnamefont{D.~H.} \bibnamefont{Youngblood}},
  \bibinfo{author}{\bibfnamefont{C.~M.} \bibnamefont{Rozsa}},
  \bibinfo{author}{\bibfnamefont{J.~M.} \bibnamefont{Moss}},
  \bibinfo{author}{\bibfnamefont{D.~R.} \bibnamefont{Brown}}, \bibnamefont{and}
  \bibinfo{author}{\bibfnamefont{J.~D.} \bibnamefont{Bronson}},
  \bibinfo{journal}{Phys. Rev. Lett.} \textbf{\bibinfo{volume}{39}},
  \bibinfo{pages}{1188} (\bibinfo{year}{1977}).

\bibitem[{\citenamefont{Lipparini and Stringari}(1989)}]{LIPPARINI1989}
\bibinfo{author}{\bibfnamefont{E.}~\bibnamefont{Lipparini}} \bibnamefont{and}
  \bibinfo{author}{\bibfnamefont{S.}~\bibnamefont{Stringari}},
  \bibinfo{journal}{Phys. Rep.} \textbf{\bibinfo{volume}{175}},
  \bibinfo{pages}{103} (\bibinfo{year}{1989}).

\bibitem[{\citenamefont{Gupta et~al.}(2016)\citenamefont{Gupta, Garg, Howard,
  Matta, {\c S}enyi{\u g}it, Itoh, Ando, Aoki, Uchiyama, Adachi
  et~al.}}]{YKGPLB2016}
\bibinfo{author}{\bibfnamefont{Y.~K.} \bibnamefont{Gupta}},
  \bibinfo{author}{\bibfnamefont{U.}~\bibnamefont{Garg}},
  \bibinfo{author}{\bibfnamefont{K.~B.} \bibnamefont{Howard}},
  \bibinfo{author}{\bibfnamefont{J.~T.} \bibnamefont{Matta}},
  \bibinfo{author}{\bibfnamefont{M.}~\bibnamefont{{\c S}enyi{\u g}it}},
  \bibinfo{author}{\bibfnamefont{M.}~\bibnamefont{Itoh}},
  \bibinfo{author}{\bibfnamefont{S.}~\bibnamefont{Ando}},
  \bibinfo{author}{\bibfnamefont{T.}~\bibnamefont{Aoki}},
  \bibinfo{author}{\bibfnamefont{A.}~\bibnamefont{Uchiyama}},
  \bibinfo{author}{\bibfnamefont{S.}~\bibnamefont{Adachi}},
  \bibnamefont{et~al.}, \bibinfo{journal}{Phys. Lett. B}
  \textbf{\bibinfo{volume}{760}}, \bibinfo{pages}{482} (\bibinfo{year}{2016}).

\bibitem[{\citenamefont{Fujiwara et~al.}(1999)\citenamefont{Fujiwara, Akimune,
  Daito, Fujimura, Fujita, Hatanaka, Ikegami, Katayama, Nagayama, Matsuoka
  et~al.}}]{Fujiwara_GR}
\bibinfo{author}{\bibfnamefont{M.}~\bibnamefont{Fujiwara}},
  \bibinfo{author}{\bibfnamefont{H.}~\bibnamefont{Akimune}},
  \bibinfo{author}{\bibfnamefont{I.}~\bibnamefont{Daito}},
  \bibinfo{author}{\bibfnamefont{H.}~\bibnamefont{Fujimura}},
  \bibinfo{author}{\bibfnamefont{Y.}~\bibnamefont{Fujita}},
  \bibinfo{author}{\bibfnamefont{K.}~\bibnamefont{Hatanaka}},
  \bibinfo{author}{\bibfnamefont{H.}~\bibnamefont{Ikegami}},
  \bibinfo{author}{\bibfnamefont{I.}~\bibnamefont{Katayama}},
  \bibinfo{author}{\bibfnamefont{K.}~\bibnamefont{Nagayama}},
  \bibinfo{author}{\bibfnamefont{N.}~\bibnamefont{Matsuoka}},
  \bibnamefont{et~al.}, \bibinfo{journal}{Nucl. Instrum. Meth. Phys. Res. A}
  \textbf{\bibinfo{volume}{422}}, \bibinfo{pages}{484} (\bibinfo{year}{1999}).

\bibitem[{\citenamefont{Brandenburg et~al.}(1987)\citenamefont{Brandenburg,
  Borghols, Drentje, Ekstr\"{o}m, Harakeh, van~der Woude, H\r{a}kanson,
  Nilsson, Olsson, Pignanelli et~al.}}]{BRANDENBURG1987}
\bibinfo{author}{\bibfnamefont{S.}~\bibnamefont{Brandenburg}},
  \bibinfo{author}{\bibfnamefont{W.~T.~A.} \bibnamefont{Borghols}},
  \bibinfo{author}{\bibfnamefont{A.~G.} \bibnamefont{Drentje}},
  \bibinfo{author}{\bibfnamefont{L.~P.} \bibnamefont{Ekstr\"{o}m}},
  \bibinfo{author}{\bibfnamefont{M.~N.} \bibnamefont{Harakeh}},
  \bibinfo{author}{\bibfnamefont{A.}~\bibnamefont{van~der Woude}},
  \bibinfo{author}{\bibfnamefont{A.}~\bibnamefont{H\r{a}kanson}},
  \bibinfo{author}{\bibfnamefont{L.}~\bibnamefont{Nilsson}},
  \bibinfo{author}{\bibfnamefont{N.}~\bibnamefont{Olsson}},
  \bibinfo{author}{\bibfnamefont{M.}~\bibnamefont{Pignanelli}},
  \bibnamefont{et~al.}, \bibinfo{journal}{Nucl. Phys. A}
  \textbf{\bibinfo{volume}{466}}, \bibinfo{pages}{29} (\bibinfo{year}{1987}).

\bibitem[{\citenamefont{Gupta et~al.}(2015)\citenamefont{Gupta, Garg, Matta,
  Patel, Peach, Hoffman, Yoshida, Itoh, Fujiwara, Hara et~al.}}]{YKG_PLB2015}
\bibinfo{author}{\bibfnamefont{Y.~K.} \bibnamefont{Gupta}},
  \bibinfo{author}{\bibfnamefont{U.}~\bibnamefont{Garg}},
  \bibinfo{author}{\bibfnamefont{J.~T.} \bibnamefont{Matta}},
  \bibinfo{author}{\bibfnamefont{D.}~\bibnamefont{Patel}},
  \bibinfo{author}{\bibfnamefont{T.}~\bibnamefont{Peach}},
  \bibinfo{author}{\bibfnamefont{J.}~\bibnamefont{Hoffman}},
  \bibinfo{author}{\bibfnamefont{K.}~\bibnamefont{Yoshida}},
  \bibinfo{author}{\bibfnamefont{M.}~\bibnamefont{Itoh}},
  \bibinfo{author}{\bibfnamefont{M.}~\bibnamefont{Fujiwara}},
  \bibinfo{author}{\bibfnamefont{K.}~\bibnamefont{Hara}}, \bibnamefont{et~al.},
  \bibinfo{journal}{Phys. Lett. B} \textbf{\bibinfo{volume}{748}},
  \bibinfo{pages}{343} (\bibinfo{year}{2015}).

\bibitem[{\citenamefont{Berman and Fultz}(1975)}]{Berman_1975}
\bibinfo{author}{\bibfnamefont{B.~L.} \bibnamefont{Berman}} \bibnamefont{and}
  \bibinfo{author}{\bibfnamefont{S.~C.} \bibnamefont{Fultz}},
  \bibinfo{journal}{Rev. Mod. Phys.} \textbf{\bibinfo{volume}{47}},
  \bibinfo{pages}{713} (\bibinfo{year}{1975}).

\bibitem[{\citenamefont{Satchler}(1987)}]{Satchler1987}
\bibinfo{author}{\bibfnamefont{G.~R.} \bibnamefont{Satchler}},
  \bibinfo{journal}{Nucl. Phys. A} \textbf{\bibinfo{volume}{472}},
  \bibinfo{pages}{215} (\bibinfo{year}{1987}).

\bibitem[{\citenamefont{Satchler and Khoa}(1997)}]{Satchler_Khoa1997}
\bibinfo{author}{\bibfnamefont{G.~R.} \bibnamefont{Satchler}} \bibnamefont{and}
  \bibinfo{author}{\bibfnamefont{D.~T.} \bibnamefont{Khoa}},
  \bibinfo{journal}{Phys. Rev. C} \textbf{\bibinfo{volume}{55}},
  \bibinfo{pages}{285} (\bibinfo{year}{1997}).

\bibitem[{\citenamefont{Rickerston}(1976)}]{DOLFIN}
\bibinfo{author}{\bibfnamefont{L.~D.} \bibnamefont{Rickerston}}
  (\bibinfo{publisher}{{unpublished}}, \bibinfo{year}{1976}).

\bibitem[{\citenamefont{Rhoades-Brown
  et~al.}(1980{\natexlab{a}})\citenamefont{Rhoades-Brown, Macfarlane, and
  Pieper}}]{ptolemy1}
\bibinfo{author}{\bibfnamefont{M.}~\bibnamefont{Rhoades-Brown}},
  \bibinfo{author}{\bibfnamefont{M.~H.} \bibnamefont{Macfarlane}},
  \bibnamefont{and} \bibinfo{author}{\bibfnamefont{S.~C.}
  \bibnamefont{Pieper}}, \bibinfo{journal}{Phys. Rev. C}
  \textbf{\bibinfo{volume}{21}}, \bibinfo{pages}{2417}
  (\bibinfo{year}{1980}{\natexlab{a}}).

\bibitem[{\citenamefont{Rhoades-Brown
  et~al.}(1980{\natexlab{b}})\citenamefont{Rhoades-Brown, Macfarlane, and
  Pieper}}]{ptolemy2}
\bibinfo{author}{\bibfnamefont{M.}~\bibnamefont{Rhoades-Brown}},
  \bibinfo{author}{\bibfnamefont{M.~H.} \bibnamefont{Macfarlane}},
  \bibnamefont{and} \bibinfo{author}{\bibfnamefont{S.~C.}
  \bibnamefont{Pieper}}, \bibinfo{journal}{Phys. Rev. C}
  \textbf{\bibinfo{volume}{21}}, \bibinfo{pages}{2436}
  (\bibinfo{year}{1980}{\natexlab{b}}).

\bibitem[{\citenamefont{Fricke et~al.}(1995)\citenamefont{Fricke, Bernhardt,
  Heilig, Schaller, Schellenberg, Shera, and \mbox{De} Jager}}]{Fricke1995}
\bibinfo{author}{\bibfnamefont{G.}~\bibnamefont{Fricke}},
  \bibinfo{author}{\bibfnamefont{C.}~\bibnamefont{Bernhardt}},
  \bibinfo{author}{\bibfnamefont{K.}~\bibnamefont{Heilig}},
  \bibinfo{author}{\bibfnamefont{L.~A.} \bibnamefont{Schaller}},
  \bibinfo{author}{\bibfnamefont{L.}~\bibnamefont{Schellenberg}},
  \bibinfo{author}{\bibfnamefont{E.~B.} \bibnamefont{Shera}}, \bibnamefont{and}
  \bibinfo{author}{\bibfnamefont{C.~W.} \bibnamefont{\mbox{De} Jager}},
  \bibinfo{journal}{At. Data and Nucl. Data Tables}
  \textbf{\bibinfo{volume}{60}}, \bibinfo{pages}{177} (\bibinfo{year}{1995}).

\bibitem[{\citenamefont{Raman et~al.}(2001)\citenamefont{Raman, Nestor, Jr.,
  and Tikkanen}}]{BE2_24Mg}
\bibinfo{author}{\bibfnamefont{S.}~\bibnamefont{Raman}},
  \bibinfo{author}{\bibfnamefont{C.~W.} \bibnamefont{Nestor}},
  \bibinfo{author}{\bibnamefont{Jr.}}, \bibnamefont{and}
  \bibinfo{author}{\bibfnamefont{P.}~\bibnamefont{Tikkanen}},
  \bibinfo{journal}{At. Data and Nucl. Data Tables}
  \textbf{\bibinfo{volume}{78}}, \bibinfo{pages}{1} (\bibinfo{year}{2001}).

\bibitem[{\citenamefont{Kib\'{e}di and Spear}(2002)}]{BE3}
\bibinfo{author}{\bibfnamefont{T.}~\bibnamefont{Kib\'{e}di}} \bibnamefont{and}
  \bibinfo{author}{\bibfnamefont{R.~H.} \bibnamefont{Spear}},
  \bibinfo{journal}{At. Data and Nucl. Data Tables}
  \textbf{\bibinfo{volume}{80}}, \bibinfo{pages}{35} (\bibinfo{year}{2002}).

\bibitem[{\citenamefont{Harakeh and Dieperink}(1981)}]{Harakeh1981}
\bibinfo{author}{\bibfnamefont{M.~N.} \bibnamefont{Harakeh}} \bibnamefont{and}
  \bibinfo{author}{\bibfnamefont{A.~E.~L.} \bibnamefont{Dieperink}},
  \bibinfo{journal}{Phys. Rev. C} \textbf{\bibinfo{volume}{23}},
  \bibinfo{pages}{2329} (\bibinfo{year}{1981}).

\bibitem[{\citenamefont{Foreman-Mackey
  et~al.}(2013)\citenamefont{Foreman-Mackey, Hogg, Lang, and
  Goodman}}]{foreman-mackey}
\bibinfo{author}{\bibfnamefont{D.}~\bibnamefont{Foreman-Mackey}},
  \bibinfo{author}{\bibfnamefont{D.}~\bibnamefont{Hogg}},
  \bibinfo{author}{\bibfnamefont{D.}~\bibnamefont{Lang}}, \bibnamefont{and}
  \bibinfo{author}{\bibfnamefont{J.}~\bibnamefont{Goodman}},
  \bibinfo{journal}{Publications of the Astronomical Society of the Pacific}
  \textbf{\bibinfo{volume}{125}}, \bibinfo{pages}{306} (\bibinfo{year}{2013}).

\bibitem[{\citenamefont{Goodman and Weare}(2010)}]{goodman-weare}
\bibinfo{author}{\bibfnamefont{J.}~\bibnamefont{Goodman}} \bibnamefont{and}
  \bibinfo{author}{\bibfnamefont{J.}~\bibnamefont{Weare}},
  \bibinfo{journal}{Communications in Applied Mathematics and Computational
  Science} \textbf{\bibinfo{volume}{5}}, \bibinfo{pages}{65}
  (\bibinfo{year}{2010}).

\bibitem[{\citenamefont{Balbutsev et~al.}(1994)\citenamefont{Balbutsev,
  Molodtsova, and Unzhakova}}]{kiev}
\bibinfo{author}{\bibfnamefont{E.~B.} \bibnamefont{Balbutsev}},
  \bibinfo{author}{\bibfnamefont{I.~V.} \bibnamefont{Molodtsova}},
  \bibnamefont{and} \bibinfo{author}{\bibfnamefont{A.~V.}
  \bibnamefont{Unzhakova}}, \bibinfo{journal}{Europhys. Lett.}
  \textbf{\bibinfo{volume}{26}}, \bibinfo{pages}{499} (\bibinfo{year}{1994}).

\bibitem[{\citenamefont{Vretenar et~al.}(2000)\citenamefont{Vretenar, Wandelt,
  and Ring}}]{Vretenar2000}
\bibinfo{author}{\bibfnamefont{D.}~\bibnamefont{Vretenar}},
  \bibinfo{author}{\bibfnamefont{A.}~\bibnamefont{Wandelt}}, \bibnamefont{and}
  \bibinfo{author}{\bibfnamefont{P.}~\bibnamefont{Ring}},
  \bibinfo{journal}{Phys. Lett. B} \textbf{\bibinfo{volume}{487}},
  \bibinfo{pages}{334} (\bibinfo{year}{2000}).

\bibitem[{\citenamefont{Kvasil et~al.}(2011)\citenamefont{Kvasil, Nesterenko,
  Kleinig, Reinhard, and Vesely}}]{Nesterenko2011}
\bibinfo{author}{\bibfnamefont{J.}~\bibnamefont{Kvasil}},
  \bibinfo{author}{\bibfnamefont{V.~O.} \bibnamefont{Nesterenko}},
  \bibinfo{author}{\bibfnamefont{W.}~\bibnamefont{Kleinig}},
  \bibinfo{author}{\bibfnamefont{P.-G.} \bibnamefont{Reinhard}},
  \bibnamefont{and} \bibinfo{author}{\bibfnamefont{P.}~\bibnamefont{Vesely}},
  \bibinfo{journal}{Phys. Rev. C} \textbf{\bibinfo{volume}{84}},
  \bibinfo{pages}{034303} (\bibinfo{year}{2011}).

\bibitem[{\citenamefont{Reinhard et~al.}(2014)\citenamefont{Reinhard,
  Nesterenko, Repko, and Kvasil}}]{Nesterenko2014}
\bibinfo{author}{\bibfnamefont{P.-G.} \bibnamefont{Reinhard}},
  \bibinfo{author}{\bibfnamefont{V.~O.} \bibnamefont{Nesterenko}},
  \bibinfo{author}{\bibfnamefont{A.}~\bibnamefont{Repko}}, \bibnamefont{and}
  \bibinfo{author}{\bibfnamefont{J.}~\bibnamefont{Kvasil}},
  \bibinfo{journal}{Phys. Rev. C} \textbf{\bibinfo{volume}{89}},
  \bibinfo{pages}{024321} (\bibinfo{year}{2014}).

\bibitem[{\citenamefont{Patel et~al.}(2013)\citenamefont{Patel, Garg, Fujiwara,
  Adachi, Akimune, Berg, Harakeh, Itoh, C.Iwamoto, Long et~al.}}]{Darshna2013}
\bibinfo{author}{\bibfnamefont{D.}~\bibnamefont{Patel}},
  \bibinfo{author}{\bibfnamefont{U.}~\bibnamefont{Garg}},
  \bibinfo{author}{\bibfnamefont{M.}~\bibnamefont{Fujiwara}},
  \bibinfo{author}{\bibfnamefont{T.}~\bibnamefont{Adachi}},
  \bibinfo{author}{\bibfnamefont{H.}~\bibnamefont{Akimune}},
  \bibinfo{author}{\bibfnamefont{G.~P.~A.} \bibnamefont{Berg}},
  \bibinfo{author}{\bibfnamefont{M.~N.} \bibnamefont{Harakeh}},
  \bibinfo{author}{\bibfnamefont{M.}~\bibnamefont{Itoh}},
  \bibinfo{author}{\bibnamefont{C.Iwamoto}},
  \bibinfo{author}{\bibfnamefont{A.}~\bibnamefont{Long}}, \bibnamefont{et~al.},
  \bibinfo{journal}{Phys. Lett. B} \textbf{\bibinfo{volume}{726}},
  \bibinfo{pages}{178} (\bibinfo{year}{2013}).

\bibitem[{\citenamefont{Youngblood et~al.}(2004)\citenamefont{Youngblood, Lui,
  Clark, John, Tokimoto, and Chen}}]{YB_116Sn_2004}
\bibinfo{author}{\bibfnamefont{D.~H.} \bibnamefont{Youngblood}},
  \bibinfo{author}{\bibfnamefont{Y.-W.} \bibnamefont{Lui}},
  \bibinfo{author}{\bibfnamefont{H.~L.} \bibnamefont{Clark}},
  \bibinfo{author}{\bibfnamefont{B.}~\bibnamefont{John}},
  \bibinfo{author}{\bibfnamefont{Y.}~\bibnamefont{Tokimoto}}, \bibnamefont{and}
  \bibinfo{author}{\bibfnamefont{X.}~\bibnamefont{Chen}},
  \bibinfo{journal}{Phys. Rev. C} \textbf{\bibinfo{volume}{69}},
  \bibinfo{pages}{034315} (\bibinfo{year}{2004}).

\bibitem[{\citenamefont{Youngblood}()}]{dhybg2}
\bibinfo{author}{\bibfnamefont{D.~H.} \bibnamefont{Youngblood}},
  \bibinfo{note}{\em private communication}.

\end{thebibliography}

\end{document}